\documentclass[aps, pra, twocolumn, superscriptaddress, reprint]{revtex4-1}

\usepackage{graphicx,amssymb,amsmath,mathrsfs,epsf,bm,cprotect,comment}
\epsfclipon

\usepackage{hyperref}

     \newcommand{\ep}{\varepsilon}
     \newcommand{\unit}[1]{\,\mathrm{#1}}

    \renewcommand{\bf}[1]{\textit{\textbf{#1}}}

     \newcommand{\expct}[1]{\langle #1 \rangle}
     \newcommand{\Expct}[1]{\left\langle #1 \right\rangle}

    \DeclareMathOperator{\sgn}{sgn}  %\UTF{0095}\UTF{0084}\UTF{008D}\UTF{0086}%
    
    \renewcommand{\eqref}[1]{Eq.~(\ref{#1})}
    \newcommand{\eqsref}[1]{Eqs.~(\ref{#1})}
    \newcommand{\pref}[1]{(\ref{#1})}
    \newcommand{\figref}[1]{Fig.~\ref{#1}}

\begin{document}

%\title{Model A dynamics in liquid crystals}
\title{Phase-ordering kinetics in the Allen-Cahn (Model A) class: universal aspects elucidated by electrically-induced transition in liquid crystals}

\author{Renan A. L. Almeida}
%\email{renan.lisboa@ufv.br}
\affiliation{Department of Physics,\! Tokyo Institute of Technology,\! 2-12-1 Ookayama,\! Meguro-ku,\! Tokyo 152-8551,\! Japan}%
\affiliation{Departmento de Física, Universidade Federal de Viçosa, 36570-900 Viçosa, MG, Brazil}
\affiliation{Department of Physics,\! The University of Tokyo,\! 7-3-1 Hongo,\! Bunkyo-ku,\! Tokyo 113-0033,\! Japan}%

\author{Kazumasa A. Takeuchi}
\email{kat@kaztake.org}
\affiliation{Department of Physics,\! The University of Tokyo,\! 7-3-1 Hongo,\! Bunkyo-ku,\! Tokyo 113-0033,\! Japan}%

%\date{\today}

\begin{abstract}
    The two-dimensional (2d) Ising model is the statistical physics textbook example for phase transitions and their kinetics. Quenched through the Curie point with Glauber rates, the late-time description of the ferromagnetic domain coarsening finds its place at the scalar sector of the Allen-Cahn (or Model A) class, which encompasses phase-ordering kinetics endowed with a nonconserved order parameter. Resisting exact results sought for theoreticians since Lifshitz's first account around a half-century ago, the central quantities of 2d Model A -- most scaling exponents and correlation functions -- remain known up to the level of competing approximations that, nevertheless, urge for substantial and accurate experimental confrontation. Here, we report a comprehensive study on the coarsening of 2d twisted nematic liquid crystals (TNLC) whose kinetics is induced by a super-fast electrical switching from a spatiotemporally chaotic (disordered) state to a two-phase concurrent, equilibrium one. Tracking the dynamics via optical microscopy, we firstly show the sharp evidence of well-established Model A aspects, such as the dynamic exponent $z=2$ and the dynamic scaling hypothesis, to then move forward and: i) confirm the Bray-Humayun theory for Porod's regime describing intradomain length scales; ii) show that Gaussian-based models, namely the Ohta-Jasnow-Kawasaki and Mazenko theories, are good descriptors of two-point spatial correlators in interdomain length scales; iii) corroborate the aging hypothesis in Model A systems which includes the collapse of two-time correlators into a master curve whose format is, actually, iv) overall best accounted for by a solution of the local scaling invariance theory: the same solution that fits the 2d nonconserved Ising model (2dIM) correlator along with the Fisher-Huse conjecture. We also v) measure a local persistence exponent with sufficient precision to distinguish between the numerically estimated 2dIM value and that exactly derived from the diffusion equation, thus suggesting the former as the true value for the Model A class. Finally, we vi) observe a fractal morphology for persistence clusters and extract their universal dimension. Due to its accuracy and possibilities, this experimental setup may work as a prototype to address further universality issues in the realm of nonequilibrium systems.
\end{abstract}

%\PACS{89.75.Da, 05.40.-a, 02.50.-r, 64.70.qj}

\maketitle

% \tableofcontents

 \section{Introduction}
 \label{Sec_Intro}

    % General view, with examples.
    Since the seminal studies \cite{Lifshitz61,Wagner61} by Lifshitz, Slyozov and Wagner on the Ostwald ripening, and those \cite{Lifshitz62,AllenCahn79} own to Lifshitz, Allen and Cahn on curvature-driven domain growth, investigations of phase ordering continue very alive as a centre of statistical physics and interdisciplinary sciences for a myriad of natural, social, and artificial phenomena \cite{Bray02,Onuki02,Puri09,Krapivsky10,Leticia15,Cates15,Berry18}.  Over the past decade, we have witnessed solid developments not only in traditional subjects, but also in the discovery on the pivotal roles that phase separation and domain coarsening play in biology and artificial active matter. The traditional road includes the unveiling of novel coarsening mechanisms such as the mechanically-driven relaxation of network-forming morphologies \cite{Tateno21}, the continued interest on spinodal decomposition of alloys and glasses \cite{Bouttes14,Geslin15}, evolving patterns of foams \cite{Lambert10,Castro14}, growing crystalline films \cite{Misbah10}, not to mention the evergreen simulations of Ising (\textit{e.g.} \cite{Arenzon15,Bhandari19,Azevedo20,Corberi20,Christiansen20}), Potts \cite{Loureiro10,Loureiro12} and voter \cite{Redner19} kinetics. The second path is directed to a quantitative understanding of emergent behaviour in active systems, which range from the intra-cellular assembly kinetics and cellular functionalization -- notably liquid-liquid phase separations involved in the formation of membraneless cell compartments -- \cite{Hyman14,Berry18,Alberti19}, to the segregation \cite{McNally17} and phase separation \cite{Liu19} within uni-cellular (bacteria) communities, pattern formation in groups of complex multi-cellular organisms \cite{Liu13,Demir20}, besides experiments \cite{Geyer19,Olivier19,Zhang21} and theoretical \cite{Cates15} accounts for artificial active matter undergoing motility-induced phase separation. Amid such a multidisciplinary environment, the iconic and perhaps the simplest example of ordering remains being the (kinetic) Ising model \cite{Glauber63,Kawasaki66} because of its simple conceptualization, theoretical and numerical tractability, both of which are accompanied by a fundamental universality that often emerge from nontrivial, real dynamics (e.g., \cite{McNally17,Wioland16}).     
    
    % Universality, MA dynamics.
    Fascination on phase-ordering kinetics then stems from its virtual ubiquitousness and complexity that yet bears universal aspects \cite{Bray02,Onuki02,Puri09,Krapivsky10,Leticia15,Cates15,Berry18}. Among these, generically one finds a dynamic scaling property that guarantees that the domain mosaic becomes statistically time-independent whenever lengths are measured in the unit of a single, asymptotically emergent, growing scale $l(t)$. Free to relax in the absence of quenched disorder, geometric frustration and external forces, domains display an algebraic growth $l(t) \sim t^{1/z}$ where $z$ is the universal dynamic exponent. For kinetics where the system's order parameter is not conserved, \textit{viz.} the Model A dynamics in the classification by Hohenberg and Halperin \cite{Hohenberg77}, one has\footnote{Exceptions to this law are the nonconserved one-dimensional XY model, which yields $z = 4$ \cite{Newman90_soc}, and the two-dimensional voter model for which clusters have no typical scale in the sense of $l(t)$, although the largest scale grows as $\sim \sqrt{t}$ \cite{Cox86,Scheucher88}.} $z = 2$ \cite{Lifshitz62,AllenCahn79,Bray_Rutenberg94}. Although simulations and experiments extensively support this dynamic scaling, rigorous derivations have been limited to the one-dimensional nonconserved kinetic Ising model (1dIM) \cite{Amar90,Bray89} and to the nonconserved $O(n\rightarrow \infty)$ model \cite{Coniglio89}, where $n$ is the number of the components of the order parameter. In the case of scalar ($n = 1$), two-dimensional (2d) continuum Model A systems quenched from a fully disordered regime to an equivalent zero-temperature ($T=0$) state, dynamic scaling with $z = 2$ is obtained from an \textit{exact} result that relies on the assumed convergence to the critical percolation fixed point \cite{Arenzon07}. Similar conclusion is reached for quenches performed from the critical state \cite{Arenzon07}, for which the cluster size distribution is exactly known \cite{Cardy03}.

    % Structure factor. OJK and TUG approximations
    Theoretically suggested \cite{Binder_combo,Furukawa79}, then observed in models \cite{Marro79} and real systems \cite{Chou81} around the end of the 70's, time-independent forms of two-point spatial correlators remain hitherto exactly known only for specific cases \cite{Amar90,Bray89,Coniglio89,Newman90_soc}. These cases do not include the scalar 2d Model A whose approximate forms were proposed along a remarkable deal of effort made around the 80's \cite{OJK82,Mazenko89} and in the 90's \cite{Mazenko90,Mazenko91_2,Bray_Humayun93,Mazenko94,Mazenko98}, where the Ohta-Jasnow-Kawasaki (OJK) theory \cite{OJK82} and the theory of unstable growth (TUG) by Mazenko \cite{Mazenko89,Mazenko90} played the major roles\footnote{Other theories such as the Kawasaki-Yalabik-Gunton (KYG) theory \cite{KYG78}, the non-Gaussian theories by Mazenko \textit{et al.} \cite{Mazenko94,Wickham97}, and the perturbative ones \cite{Bray_Humayun93,Mazenko98}, generically recover either OJK or TUG in the leading order. The KYG theory recovers the OJK function for a particular value of OJK diffusion constant \cite{Bray02}.}. Tested against discrete \cite{Humayun92,Blundell93} and continuum \cite{Bray93,Brown98} models, both OJK and TUG functional forms generically succeeded\footnote{
    %See a specific counterexample in \cite{Blundell93}.
    However, Ref.\cite{Blundell93} pointed out that Gaussian theories such as OJK and TUG do not describe, simultaneously, data of the correlator of the order parameter field and that of the squared field. Limitations of Gaussian approximations were also discussed in Ref.\cite{Yeung94,Yeung18}.
    } in fitting to data. The OJK form was tested in an notable experiment \cite{Orihara86} back to the 80's, although agreement was mostly confined to the short intradomain scale ruled by the Porod's regime \cite{Porod51_combo}.

    % Bray exact results for the Porods scale.
    On the Porod's regime, Bray and Humayun \cite{Bray_Humayun93_amplitudes} derived exact formulae for the short-distance limit of the spatial correlator (or the tail of the structure factor) -- hereinafter referred to as the Bray-Humayun amplitude --, so that there is no need to rely on approximate theories in this regime. Moreover, because the formulae were derived under minimal assumptions on the existence of defects, the Bray-Humayun amplitude shall be valid regardless of the conservation laws of the order parameter and whether the dynamic scaling holds \cite{Bray_Humayun93_amplitudes}. Despite so, to our knowledge, no experimental measurement has turned attention to this prediction.
    
    % Autocorrelation exponent
    The memory of systems in the coarsening stages is an additional aspect of interest because corresponding two-time and infinitely-many time statistics may also display scale invariance \cite{Bray02,Majumdar99_rev,Bray13_rev}, including aging akin to that observed in glassy materials \cite{Henkel}. As revealed by Fisher and Huse \cite{Fisher88}, the description of two-time correlators requires an additional nonequilibrium exponent -- defined in Sec.~\ref{Sec_Time_subsec_ageing}, \eqref{eq-time-covscaling} -- $\lambda(n,d)$ \cite{Fisher88,Newman90_On_combo} which is affected by the presence of long-range order in initial conditions \cite{Humayun91,Corberi16}. For the Model A class, exact results for $\lambda$ are known for the 1dIM \cite{Bray89}, the 1d noiseless time-dependent Ginzburg-Landau (TDGL) equation \cite{Bray95}, including for the $O(n)$ model in terms of the $1/n$ expansion \cite{Newman90_On_combo}. In the scalar, 2d case, Fisher and Huse conjectured $\lambda_{\textrm{FH}} = 5/4$ based on the convergence of the dynamics to the critical percolation fixed point, which then supplements their heuristic arguments \cite{Fisher88}. Such conjecture was initially supported by 2dIM \cite{Fisher88,Humayun91,Corberi16} and 2dTDGL simulations \cite{Brown97}, but recent studies based on finite-size scaling for the very same models suggest a value closer to the prediction by TUG \cite{Midya14,Vadakkayil18}, $\lambda_{\textrm{TUG}} \approx 1.29$ \cite{Liu91}. In parallel, the OJK theory predicts a rather different value,  $\lambda_{\textrm{OJK}} = 1$ \cite{Yeung90}, coinciding with the lower bound of the inequality $\lambda \geq d/2$ for nonconserved dynamics \cite{Fisher88,Yeung96}. The OJK value has not been observed so far but for 2dIM simulations with strongly long-range interactions \cite{Christiansen20}. Noting that deviations from the $\mathbb{Z}_2$-symmetric dynamics \textit{must} be taken into account when analysing the two-time correlator -- otherwise strongly biasing and artificially changing its corresponding asymptotic decay -- it is important to elucidate on the experimental support to $\lambda_{\textrm{FH}}$ in Ref. \cite{Mason93} in the light of dealing with systems that are not $\mathbb{Z}_2$-symmetric, as it happens in experiments. 
    %. Given this situation, it is important to have sufficiently precise experimental measurement of $\lambda$ to distinguish the different predictions, taking also into account possible effects of deviations from the ideal setting, such as bias in the magnetization.
    % nor the neglection of possible effects of a nonvanishing magnetization \cite{Mason93} dynamics on the estimate of $\lambda$, shall be taken as settled.
            
    % Functional form
    More fundamentally, experimental support to the dynamic scaling according to aging hypothesis \cite{Henkel} is surprisingly elusive for the Model A class. As consequence, functional forms for the two-time correlator given by OJK \cite{Yeung90}, TUG \cite{Liu91}, and the local scale invariance (LSI) theory by Henkel \cite{Henkel02}, all lack confrontation with real dynamics.
    
    % Persistence
    Another nontrivial facet of phase-ordering kinetics is encoded in their first-passage statistics \cite{Redner01_book}, specifically in the local persistence probability $Q(t,t_0)$ \cite{Majumdar99_rev,Bray13_rev,Aurzada15_rev}. In the language of the Ising model, $Q(t,t_0)$ is defined as the fraction of spins at time $t$ which has never flipped since a previous time $t_0$.
    By construction, it involves correlation of infinitely-many points. In the thermodynamic limit ($L \rightarrow \infty$, followed by $t \rightarrow \infty$ with fixed $t/t_0$), $Q$ often decays as $Q(t,t_0) \sim (t/t_0)^{-\theta}$ for large $t/t_0$ with the persistence exponent $\theta$, which is normally independent of $z$ and $\lambda$ \cite{Majumdar99_rev,Bray13_rev}. The associated body of knowledge was seeded in the numerical work by Derrida \textit{et al.} \cite{Derrida94} on the 1d $q$-state Potts model (1dPM), then grew and took shape thanks to 1d-5d IM simulations \cite{Stauffer94}, analytical approaches for the noiseless 1d TDGL equation \cite{Bray95}, reaction-diffusion schemes \cite{Krapivsky94,Cardy95}, exact solutions for 1dPM \cite{Derrida95,Derrida96}, approximations for simple diffusion (DIF) with random initial conditions (RIC) \cite{Majumdar96_diff,Derrida96_OJK}, besides being nurtured by experiments \cite{Marcos95,Tam97,Yurke97,Wong01,Soriano09}. Solvability of $Q$ in Model A systems is constrained to 1d cases \cite{Bray95,Derrida95,Derrida96}. In the scalar sector of 2d Model A, early numerics predicted $\theta \approx 0.22$ \cite{Derrida94,Stauffer94} while a recent 2dIM simulation suggests $\theta = 0.199(2)$ \cite{Blanchard14}. 
    %This agrees with an approximation result $\theta \approx 0.19$ by Majumdar and Sire \cite{Majumdar96_higherd}, obtained by combining the same Gaussian approximation as TUG used and the time correlator from 2dIM simulations. 
    This value is rather discernible from the exact $\theta_{\textrm{DIF}} = 3/16 = 0.1875$ value derived from the 2d DIF-RIC \cite{Poplavskyi18} and OJK theory \cite{Derrida96_OJK}, which is also faced as a candidate value for 2d Model A. The early experiment by Yurke, Pergellis, Majumdar and Sire reported $\theta = 0.19(3)$ \cite{Yurke97}, albeit precision does not allow to rule out $\theta_{\textrm{DIF}}$; in this particular case, one requires a more accurate outcome. Furthermore, the fractal morphology of clusters formed by local persistence spins in 2dIM \cite{Manoj00_pre,Jain00} calls for substantiation.
    
    A traditional experimental approach to nonconserved kinetics has been the twisted nematic liquid crystal (TNLC) \cite{Orihara86,Mason93,Yurke97}, which can be quenched through a first-order transition between the high-temperature fluid phase and the lower-temperature nematic phase. Not forgetting to pay due tribute to these previous systems that aimed at measuring dynamic scaling properties, we however note that the inexorable slow quench rate implied in the isotropic-nematic transition and the issues of metastability, nucleation and nuclei growth stages, have conspired to blur comparisons with predictions made from genuinely sudden transitions across second-order points. It is also highly desirable to seek for a versitle setup in which different sorts of kinetics can be triggered.
    
    Here, we exploit electrically-driven regimes in TNLC layers to induce a virtually instantaneous switch from a spatiotemporally chaotic (disordered-like) state to an equilibrium, two-phase competing state. Because of the sudden removal of the driving, problems related to first-order transitions are negligible (if existent), while thermal effects are finely controlled. In addition, this setup offers an elegant way to induce different sorts of kinetics by transiting between nonequilibrium states. We focus on the simplest transition type that is towards the equilibrium, so that we assess the several aspects on 2d Model A waiting for experimental elucidation. 
    
    We firstly show well-established Model A features such as the sharp evidence of dynamic scaling with $z = 2$. Then we move to assess the Bray-Humayun amplitude and test OJK and TUG theories in length scales much larger than the Porod's regime. Paying attention to the asymmetry in the twisted phases, an aging hypothesis for Model A systems is confirmed by a rescaled two-time correlator whose master form is used to confront the OJK, TUG and LSI theories. %We also observe dynamic (aging) scaling of the two-time correlator, paying attention to asymmetry between the two TNLC phases which may affect the result. We obtain the scaling function for the two-time correlator and compare it with the OJK, TUG, and LSI predictions. 
    We find that a particular solution of LSI, which fits 2dIM data and assumes the Fisher-Huse conjecture, gives the superior account for the experiment. Although our estimate of $\lambda$, in the asymptotic regime, agrees with both the FH and TUG values within the uncertainty, analysis from the correlator form itself rules out the TUG scenario from a global viewpoint. The local persistence statistics is also investigated; we measure a precise value of $\theta$ that agrees with the numerically estimated value for 2dIM, but that declines the suggestion of the analytic DIF-RIC (OJK) theory. Morphology of persistence clusters is shown as a fractal whose dimension is in harmony with that estimated from 2dIM.
    
    % Way of presentation
    This contribution is organized as follows. Section~\ref{Sec_Materials} introduces the electrohydrodynamic phenomenon of nematic liquid crystals, the experimental setup and the protocol used to trigger the ordering kinetics. Morphology of TNLC domains and the shrinking rate of their interfaces are the opening themes of Sec.~\ref{Sec_Space}. The sequence quantifies the asymmetry in twisted phases. Two-point spatial correlators, the growth law, dynamic scaling and the Bray-Humayun amplitude are addressed in Sec.~\ref{Sec_Space_subsec_covariance}. Comparison with OJK and TUG functions closes Sec.~\ref{Sec_Space}. Time correlation properties are studied in Sec.~\ref{Sec_Time}. Two-time correlators are presented in Sec.~\ref{Sec_Time_subsec_ageing}; it covers the measurement of $\lambda$, dynamic scaling aspects, and tests of theoretical forms. Local persistence probability and morphology of persistence clusters are presented in Sec.~\ref{Sec_Time_subsec_persistence}. We conclude this contribution in Sec.~\ref{Sec_Conclusions}.

 %--------------------------------------------------------------
 %\section{Material and methods} 
 \section{Experimental setup and electrical switching}
 \label{Sec_Materials}
 
We exploit the electrohydrodynamic convection of nematic liquid crystal \cite{Williams63,Carr69,Helfrich69,Orsay72,Ribotta86} that arises when a thin layer (typically 10 to 100$\unit{\mu m}$) of nematic liquid crystal -- with negative dielectric anisotropy ($\ep_\parallel - \ep_\bot < 0$)\footnote{By convention, quantities with $\parallel$ and $\bot$ subscripts indicate the component parallel and perpendicular to the director field, respectively.} and positive conductance anisotropy ($\sigma_\parallel - \sigma_\bot > 0$) -- is subjected to an electric voltage of $V \sim 10\unit{V}$ or higher \cite{DeGennes93}. Upon increasing $V$, one observes a sequence of convective patterns \cite{Joetz86,Joetz86_2,Kai89} and finally a spatiotemporally chaotic state called the dynamic scattering mode 2 (DSM2)\footnote{There also exists a regime called dynamic scattering mode 1 (DSM1) that is observed at lower applied voltages. DSM1 is characterized by the absence of sustained elongated disclinations and by the presence of anisotropy along the direction of the underlying roll structure \cite{Krupowski78,Nagaya99,Strangi99}.}\cite{Heilmeier68,Sussman72,Nehring72,Chang73}. DSM2 stands out by the presence of a high density of disclinations; fluctuations of the director field are short-range correlated in space and time: $\approx$ 1-3$\unit{\mu m}$ and 10$\unit{ms}$, respectively, for a 50$\unit{\mu m}$ thick layer \cite{Joetz86}. We use DSM2 as a disordered initial condition to study relaxation towards the equilibrium state.

\begin{figure*}[!t]
        \centering
        \includegraphics[width=0.99\hsize,clip]{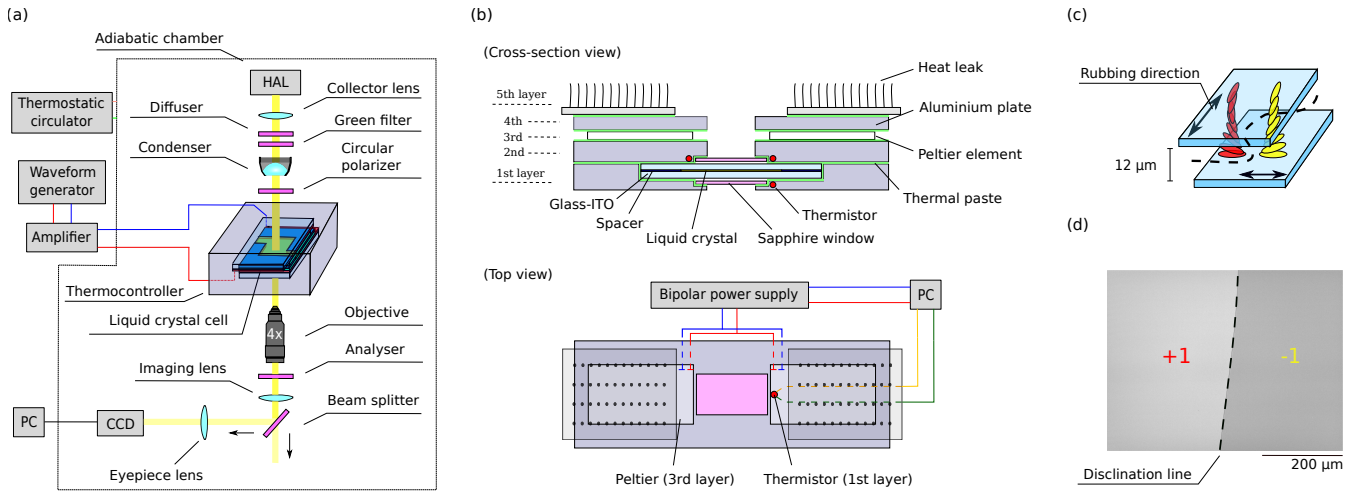}
        \caption{(a) Schematics of the experimental setup. (b) Sketch of the thermocontroller. Using a personal computer (PC), we convert the resistance of a thermistor to the temperature and determine the current to drive the Peltier elements by a proportional-integral-derivative (PID) operation. (c) Illustration of a TNLC cell in which the director field is represented by coloured ellipses; different colours represent the distinct macroscopic twist orientations. Dashed line represents a domain boundary (disclination). (d) Part of a real TNLC image with domains of opposite handedness detected by bright ($\phi = +1$) and dark ($-1$) shades of grey. A dashed line is drawn (not in scale for the sake of visualization) along the domain boundary. Acronyms stand for: halogen lamp (HAL), charged-coupled camera (CCD), personal computer (PC). Objects are not to scale in panels (a)-(c).}
        \label{Fig_Material-instrument}
        \end{figure*}

We built a capacitor of parallel glass plates coated with indium tin oxide. The gap was set by polyester spacers of thickness $12\unit{\mu m}$ and enclosed an empty region $16\unit{mm} \times 16\unit{mm}$. Prior to assembly, a polyvinyl alcohol film was coated over each plate and rubbed to set the direction of the surface alignment. Rubbing directions were set to be orthogonal between the two plates to realize a twisted nematic liquid crystal (TNLC) cell \cite{Mauguin1911}. The capacitor was then filled with $N$-4-methoxybenzylidene-4-butylaniline (purity $> 98.0\%$, Tokyo Chem. Ind., Japan) doped with $0.01\unit{wt\%}$ of tetrabutylammonium bromide (Tokyo Chem. Ind., Japan). Director field twisted along the bulk in either left- or right-handed orientation in the cell [\figref{Fig_Material-instrument}(c)].

    The cell was inserted in a handmade thermocontroller [\figref{Fig_Material-instrument}(b)]; the set was placed on the stage of an inverted microscope (IX73, Olympus, Japan). Green-filtered, circularly polarised light was emitted through the cell; transmitted images were recorded by a charge-coupled device [\figref{Fig_Material-instrument}(a)]. We monitored an area of dimensions $L_x \times L_y = 2.9 \unit{mm} \times 2.2\unit{mm}$ in the sample by a $\times$4 objective. The whole apparatus was inserted in a thermally isolated chamber whose inner temperature was kept roughly constant by a thermostat circulator. 
    %Temperature of the cell was maintained by the thermocontroller. 
    Peltier elements in the thermocontroller were connected to a bipolar current supply set by a proportional-integral-derivative feedback using the signal of one of the thermistors located at $1\unit{mm}$ to the cell. During the experiments, we kept the cell temperature at $25\unit{^\circ C}$ with fluctuations smaller than $\pm 1.5 \unit{mK}$ according to the reference thermistor. 
    %Prior to the experiments, we also evaluated the degree of in-plane, transversal and diagonal temperature fluctuations, which were less than 1$\unit{mK}$, 40$\unit{mK}$, and 50$\unit{mK}$, respectively. 
    Prior to the experiments, we also monitored the temperature at different positions near the cell. Thermal fluctuations of thermistors not used for the feedback control were at most $\pm 10\unit{mK}$, whilst the temperature gradient formed across the cell was a few tens of mK.
 
    Ordering kinetics was triggered as follows. We applied a 100$\unit{Hz}$, 70$\unit{V}$ sinusoidal voltage across the cell to generate the DSM2 state.
    %Using an external wave generator and an amplifier, we applied an electric field to the TNLC layer. We applied a 100$\unit{Hz}$, 70$\unit{V}$ sinusoidal voltage to generate the DSM2 state. 
    Equilibrium twist alignment of the director field was then replaced by disordered and fluctuating configurations. The cell was kept in DSM2 during 120$\unit{s}$ ($\sim 10^4$ correlation times of the DSM2 state) before we instantaneously\footnote{The time taken for this removal of the electric field was much shorter than $10\unit{ms}$, which was the period of the alternating field and also the order of correlation time of the DSM2 state \cite{Joetz86}.} removed the electric field. From this switching time $t = 0\unit{s}$ onwards, the twisted boundary conditions for the director field induced either left- or right-handed twist; these twisted alignments were formed after complex disentangling of disclination loops. We recorded the dynamics at 3$\unit{Hz}$ over $2000\unit{s}$ with pixel size $a \approx 1.82\unit{\mu m}$; 20 independent relaxation histories were collected. Due to the polarization optics, twists of opposite handedness were distinguished in the images by darker and brighter shades of grey [\figref{Fig_Material-instrument}(d)]. We attributed to each pixel a local state variable $\phi(\bm{r},t)$ that took $\phi = - 1$ and $+ 1$ at the dark- and bright-grey colour pixels, respectively. 
    Domains of the opposite handedness were bordered by disclination lines [\figref{Fig_Material-instrument}(c) and (d)] that appeared as dark lines of thickness $\xi \approx a$. Because it was hard to fully detect disclinations out of dark-grey colour pixels, we regarded disclinations as part of the $\phi = - 1$ phase: a fair approximation because $\xi \approx a$ and because interest is in the scaling regime where $\xi/l(t) \rightarrow 0$. Nonetheless, we warn for specific instances where the binarization effect might play some role. Data for $t \leq 2.33\unit{s}$ are neglected because the large fraction of disclinations blurred rational detection of domains.
    %Because of this, we analyzed images after $t = 2.4\unit{s}$, for which the disclination density was negligible ($\xi \ll l(t)$) and we can justify the binary representation we adopted. Notwithstanding, we explicitly make warnings in the text for specific cases in which binarization effects may play a role.

 %--------------------------------------------------------------
 \section{Spatial scaling and phase asymmetry}
 \label{Sec_Space}

    \subsection{Shrinkage law}
    \label{Sec_Space_subsec_shrink}

        \begin{figure*}[!t]
        \centering
        \includegraphics[width=0.99\hsize,clip]{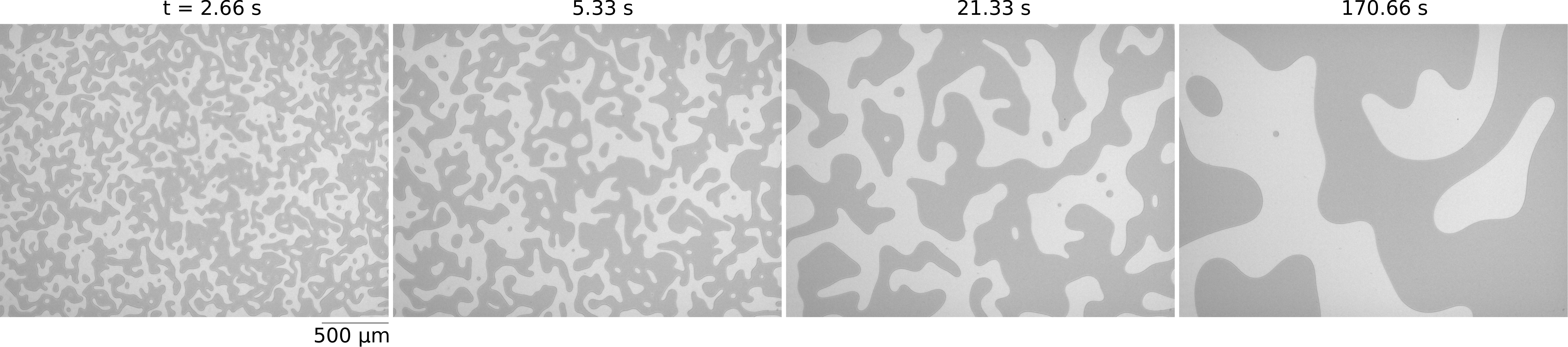}
        \caption{Configurations of the ordering of twisted nematic liquid crystals (TNLC) during a relaxation towards the equilibrium. The dynamics was triggered by a virtually instantaneous electrical switch (at $t = 0\unit{s}$) from a spatiotemporally chaotic called dynamic scattering mode 2 (DSM2) towards the two-phase competing, equilibrium state. Different shades of grey correspond to twists of opposite handedness in the nematic director field. See also Supplemental Videos 1 and 2.}
        \label{Fig_Space-domains}
        \end{figure*}

    Figure \ref{Fig_Space-domains} shows TNLC domain configurations in ordering (see also Supplemental Videos 1 and 2). Configurations at $t = 2.66\unit{s}$ appear as  a random patchwork of hundreds of domains (or clusters) that progressively coarsen afterwards. Initial irregular contours of domains become smoother in time. While the total length of boundaries tends to decrease, larger domains increase their area at the expense of smaller shrinking ones. Around ten clusters are left at $t \approx 150\unit{s}$, but ordering is only completed a decade later, at a finite-size equilibration time $t^* \approx 1.2 \times 10^{3}\unit{s}$.

        \begin{figure}[!t]
        \centering
        \includegraphics[width=6.5cm]{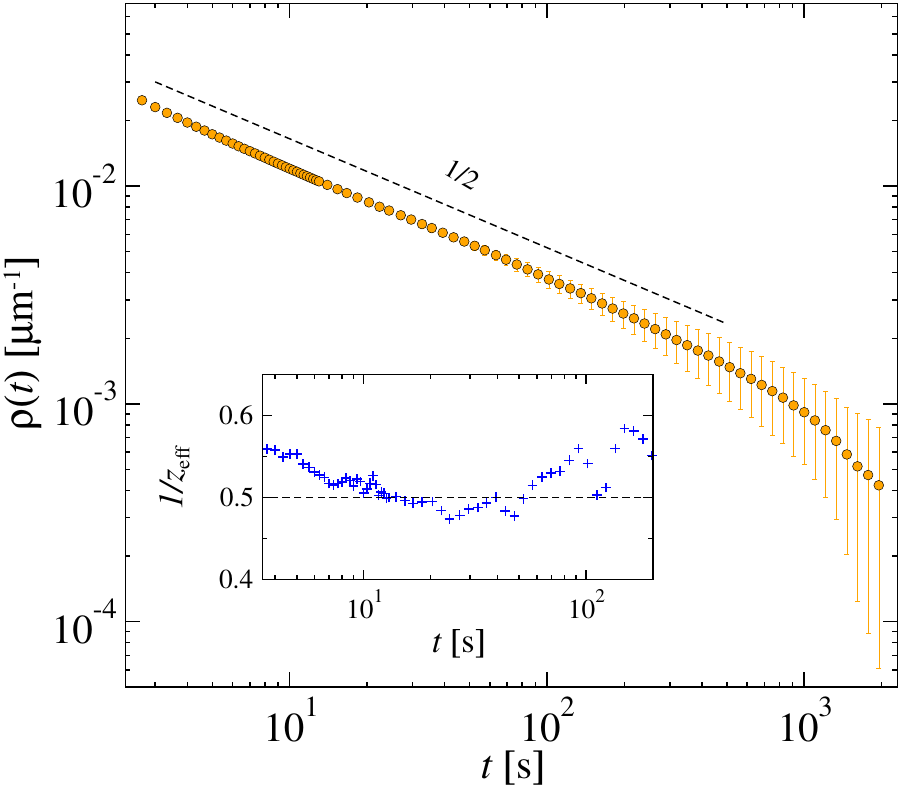}
        \caption{Density of the interface length in time. Error bars indicate one standard deviation. Dashed black line is a guide for eyes with the slope $-1/2$ expected for Model A dynamics. Inset shows the time series of the effective exponent, $1/z_{\textrm{eff}} = - d(\ln\rho(t))/d\ln t$, which fluctuates around $1/2$ (dashed black line) within the core of the power-law decay, $\rho(t) \sim t^{-1/2}$, for $6\unit{s} \lesssim t \lesssim 60\unit{s}$. }
        \label{Fig_Space-interface}
        \end{figure}

    To quantify shrinkage of interfaces (disclinations), we measure the interface density $\rho(t)$ as the number of neighbouring pairs with the opposite handedness
    %[\textit{$\phi(\textbf{r},t)$, $\phi(\textbf{r} + \textbf{a},t)$}] with configuration [$\pm 1, \mp 1$],
    multiplied by $a/L_x L_y$. Based on scaling hypothesis \cite{Bray02}:
    
        \begin{equation}
        \rho(t) \sim t^{-1/z}.
        \label{eq-space-interfacedensity}
        \end{equation}
    
    Figure~\ref{Fig_Space-interface} shows that such a decay, dictated by a slope near $1/2$,  occurs in the experiment for nearly two decades. The decay is better quantified by the effective exponent, $z^{-1}_{\textrm{eff}}(t) = - d(\ln\rho(t))/d\ln t $, whose plateau for $6\unit{s} \lesssim t \lesssim 60\unit{s}$ indicates the core of the algebraic regime (\figref{Fig_Space-interface}, inset). Time averaging in this regime, or fitting \eqref{eq-space-interfacedensity} to the data of each realization and averaging, we obtain:
    
        \begin{equation}
        1/z = \begin{cases}
        0.505(15)  & \mbox{(effective exponent);} \\
        0.498(6) & \mbox{(power-law fit).}
        \end{cases}
        \label{eq-space-1z}
        \end{equation}
    The number(s) into parentheses refers to the uncertainty (standard deviation for the effective exponent; standard error for power-law fit) in the last digit(s). Converting to $z$: $z = 1.98(6)$ (effective exponent) and $z = 2.008(22)$ (power-law fit). These results are in accurate agreement with the theory for curvature-driven interface dynamics \cite{Lifshitz62,AllenCahn79,Bray_Rutenberg94}, a hallmark of the Model A systems.
 
        \begin{figure*}[!t]
        \centering
        \includegraphics[width=6.5cm]{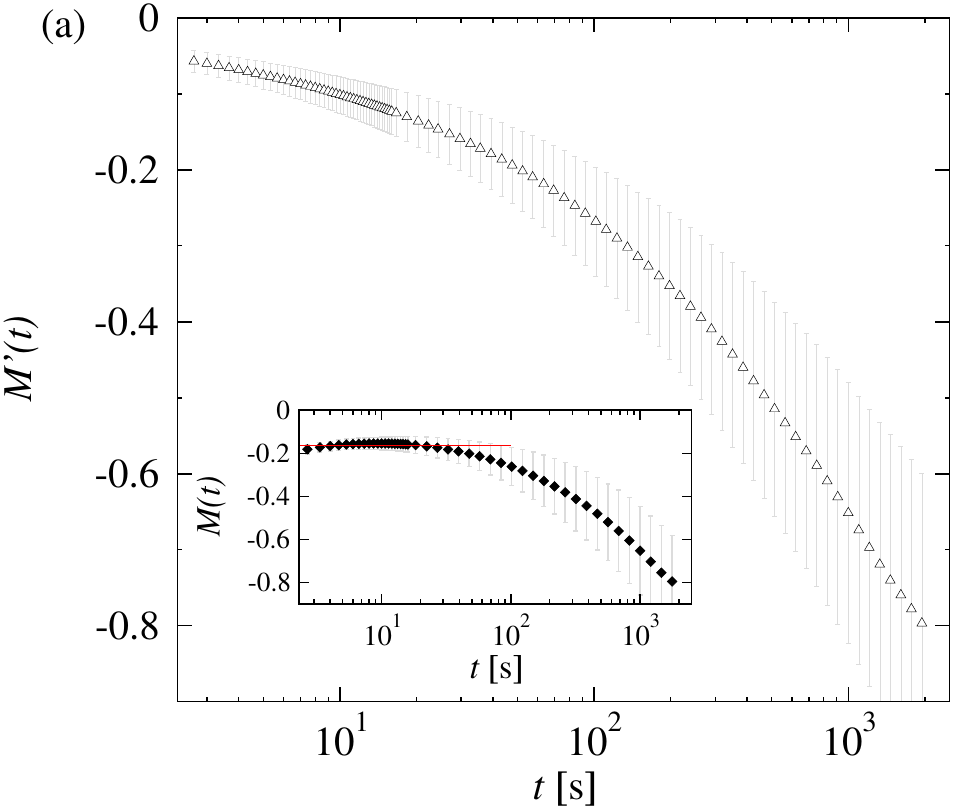} \quad
        \includegraphics[width=6.5cm]{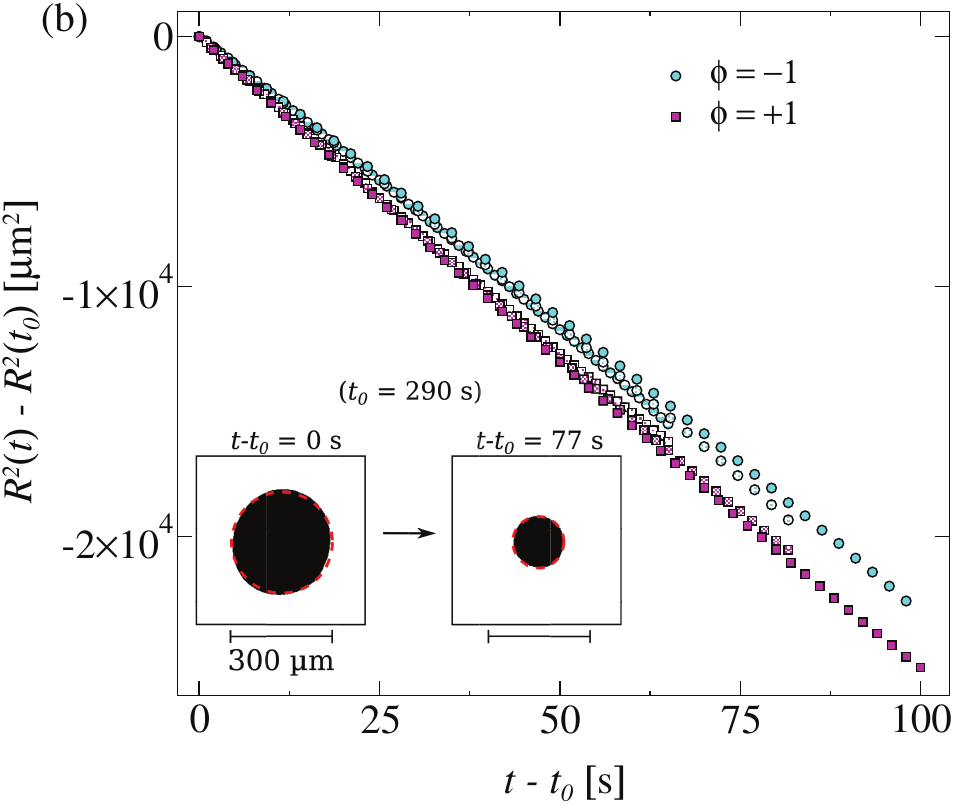}
        \caption{(a) ``Magnetization'' of the TNLC system in time with no contribution of disclinations (free from binarization effects; see text). The inset shows the bare magnetization, $M(t)$; the guide line (red) marks the averaged value $\overline{M} \approx -0.17$ for $2.66\unit{s} \leq t \leq 60\unit{s}$. In both the panels, error bars indicate one standard deviation. (b) Evolution of the square radius $R^2(t)$ (converted from the area) of approximately circular TNLC domains of phase $\phi$ surrounded by a ``sea'' of the opposite phase. Curves with different patterns and shades of cyan circles (magenta squares) are independent realizations for bubbles made of the slightly favoured (unfavoured) $\phi = -1$ ($+1$) phase. The insets show snapshots of a $\phi = -1$ bubble at two different times, where the red dashed lines is a guide for a circle of radius $R(t)$ at each time.}
        \label{Fig_Space-bubble}
        \end{figure*}

    \subsection{Phase asymmetry and domain shrinkage}
    \label{Sec_Space_subsec_assymetry}
 
    The twist phases in the cell are not strictly $\mathbb{Z}_2$-symmetric because the rubbing method inserts a nonvanishing pretilt angle for the director field at the anchoring surfaces \cite{Becker86}. Misalignment between top and bottom surfaces may add for the symmetry breaking \cite{Nagaya87}. 
    
    Asymmetry in the kinetics is captured by the ``magnetization'', $M(t) = 1 - 2A(t)$, defined from the area fraction $A(t)$ of the $\phi = -1$ (darker) phase [Fig.~\ref{Fig_Space-bubble}(a), inset]. Note that $M(t)$ contains a contribution from disclinations included in $\phi = -1$. To remove this effect, we carried out an extra set of 20 independent experimental runs with the cell placed upside down: by this changing, the brightness of the two competing domains is inverted, so that the sign of $\phi$ was flipped -- except for disclinations which, keeping looking as dark-coloured pixels, remained regarded as $\phi = -1$ after binarization. Measuring the magnetization $M_\mathrm{flip}(t)$ for this sign-flipped data set, we compute $M'(t) = [M(t) - M_\mathrm{flip}(t)]/2$ as the disclination-free $M'(t)$ magnetization [Fig.~\ref{Fig_Space-bubble}(a), main panel]. The $M'(t)$ reveals that the $\phi = -1$ phase is slightly favoured in the main data set, albeit asymmetry is small; for instance, $|M(t)|, |M'(t)| \lesssim 0.2$ for $t \lesssim 60\unit{s}$. Time average in this interval gives $\overline{M} = -0.17(3)$, which corresponds to $\overline{A} = 0.58(2)$.
    %The time averages of $|M(t)|$ and $|M'(t)|$ during this time window are $\overline{M} = -0.17(3)$ and $\overline{M'} = -0.11(3)$, respectively (the numbers in the parentheses indicate the time-averaged standard deviation\footnote{More precisely, the standard deviation of $M(t)$ over realizations was computed for each $t$ and its time average is shown in the parentheses for the estimate of $\overline{M}$. For $\overline{M'}$, we evaluated the variation of $M'(t)$ from the standard deviation of $M(t)$ and $M_\mathrm{flip}(t)$ and took the time average.}). In terms of the area fraction, those are $\overline{A} = 0.583(17)$ and $\overline{A'} = 0.557(14)$. Similar degree of asymmetry, $\overline{A} = 0.576(16)$, was reported in an earlier study \cite{Yurke97}. 
    %We shall therefore use this time window preferentially in the following analyses.
    %For $t \leq 60\unit{s}$, the time average of $|M(t)|$ was $\overline{M} = -0.17(3)$ (the number in the parentheses indicates the time-averaged standard deviation).
    %In terms of the area fraction, this corresponds to $\overline{A} = 0.58(2)$, which is comparable to $\overline{A} = 0.576(16)$ reported in an earlier study \cite{Yurke97}.
    
    Asymmetry is also seen in the shrinkage rate of domains of phase $\phi$ surrounded by a ``sea'' of $-\phi$. For curvature-driven dynamics, the squared radius $R^2(t)$ of a circular domain (or ``bubble'') evolves, from a reference time $t_0$, as \cite{AllenCahn79}:

        \begin{equation}
        R^2(t) - R^2(t_0) = -2D_{\textrm{AC}}(t - t_0) + K,
        \label{eq-space-bubble}
        \end{equation}
    where $D_{\textrm{AC}}$ is the Allen-Cahn diffusion parameter and $K$ is a constant. We monitored 6 independent, isolated and nearly circular bubbles; 3 of them made of the $\phi = - 1$ favoured phase, and the remaining ones made of $\phi = + 1$. Measuring their area and equalizing $\pi R^2(t)$ to define $R^2(t)$ for each bubble, we accompany their squared radius in time [\figref{Fig_Space-bubble}(b)]. Fitting according \eqref{eq-space-bubble} yields $D_{\textrm{AC}}^{-} = 117.3(17)\unit{\mu m^2s^{-1}}$ and $D_{\textrm{AC}}^{+} = 126.0(6)\unit{\mu m^2s^{-1}}$ for $\phi = - 1$ and $+1$ bubbles, respectively. The asymmetry translates in nearly $4\%$ of deviation of each diffusion constant relative to the average,
         \begin{equation}
        D_{\textrm{AC}} = 122(4)\unit{\mu m^2s^{-1}},
        \label{eq-space-DacTNLC}
        \end{equation}
    where the number in the parentheses was set to cover the interval $[D_{\textrm{AC}}^{-},D_{\textrm{AC}}^{+}]$.
 
    %Unlike previous liquid crystal systems \cite{Orihara86,Mason93,Yurke97,Sicilia_exp08}, the initial state of the experiment is the spatiotemporally chaotic DSM2 state.
    We remind that the initial state of the experiment is DSM2, which may provide an unbiased (or weakly biased) initial condition because the surface anchoring of the director field is not preserved \cite{Fazio99} and because the bulk director is strongly disturbed by the high density of disclinations. Thus, the observed asymmetry -- including the one in the earlier times not captured in Fig. 4(a) -- is likely developed along with the kinetics instead of being set by initial conditions. 
    
    %We therefore consider that the observed asymmetry is mainly developed along with the kinetics [including that leading to initial formation of twist domains, which is not characterized in \figref{Fig_Space-bubble}(a)].

    %Experiments with asymmetries $|\overline{M}| \lesssim 0.3$ have been used \cite{Orihara86,Mason93,Yurke97,Sicilia_exp08} to test theoretical predictions of $\mathbb{Z}_2$-symmetric models \cite{Bray02}, but the presence of asymmetry has not been taken into account in the analysis. As we shall see, compensating the influence of the asymmetry is a simple task for the spatial correlator addressed in Sec.~\ref{Sec_Space_subsec_covariance}, but the situation is a little more involved for the two-time correlator discussed in Sec.~\ref{Sec_Time_subsec_ageing}. Since the weak violation of the $\mathbb{Z}_2$ symmetry is an issue often encountered in experiments, we believe this point deserves due attention.

    The usual presence of asymmetry in experiments demands additional care when  comparison is made with predictions for $\mathbb{Z}_2$-symmetric models. As we shall see, although this a simple task for some quantities such as the spatial correlator (Sec.~\ref{Sec_Space_subsec_covariance}), they are subtle for, e.g., the time correlator (Sec.~\ref{Sec_Space_subsec_covariance}).
    %Since the weak violation of the $\mathbb{Z}_2$ symmetry is the rule in experiments \cite{Orihara86,Mason93,Sicilia_exp08}, yet not taken into account in the analysis, we believe this point deserves due attention.
    
        \begin{figure*}[!t]
        \centering
        \includegraphics[width=6.5cm]{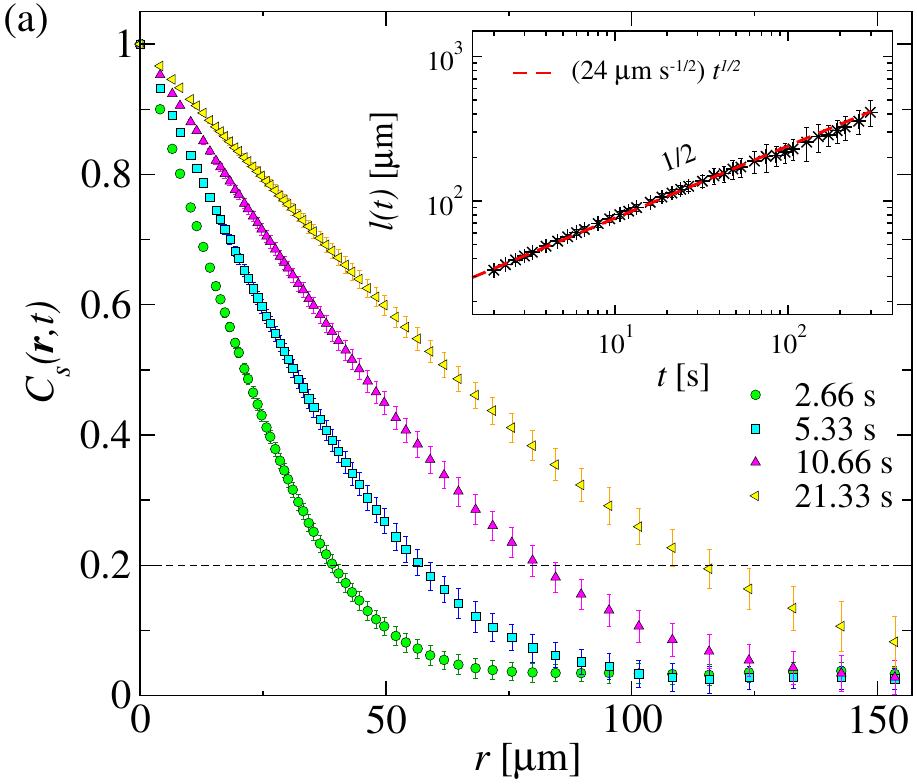} \qquad
        \includegraphics[width=6.5cm]{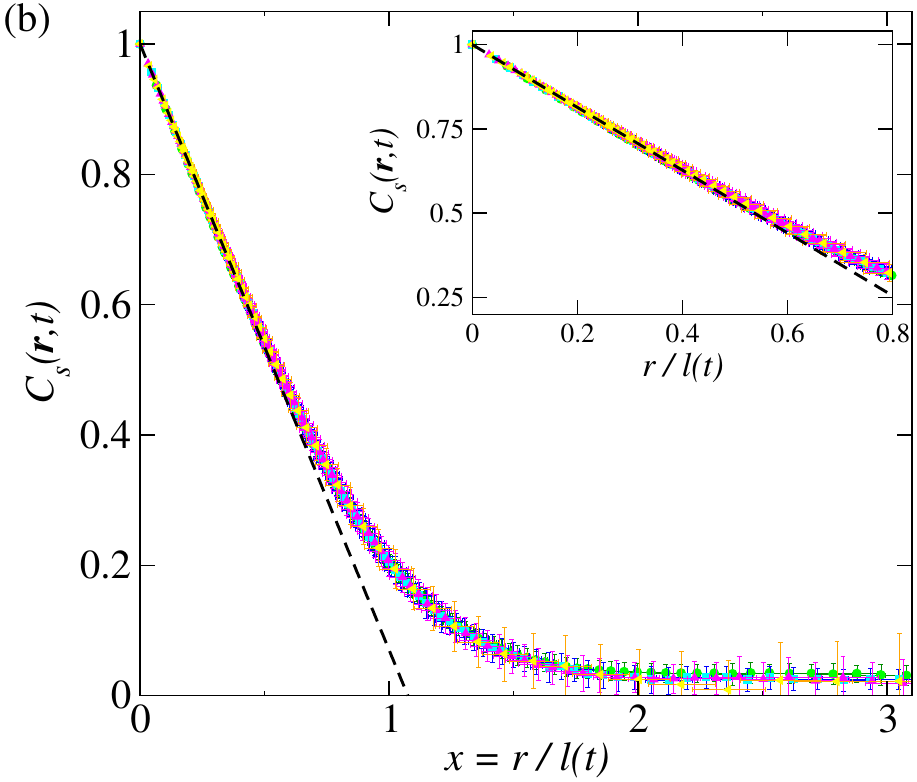} \qquad
 
            \vspace{0.5cm}
 
        \includegraphics[width=6.5cm]{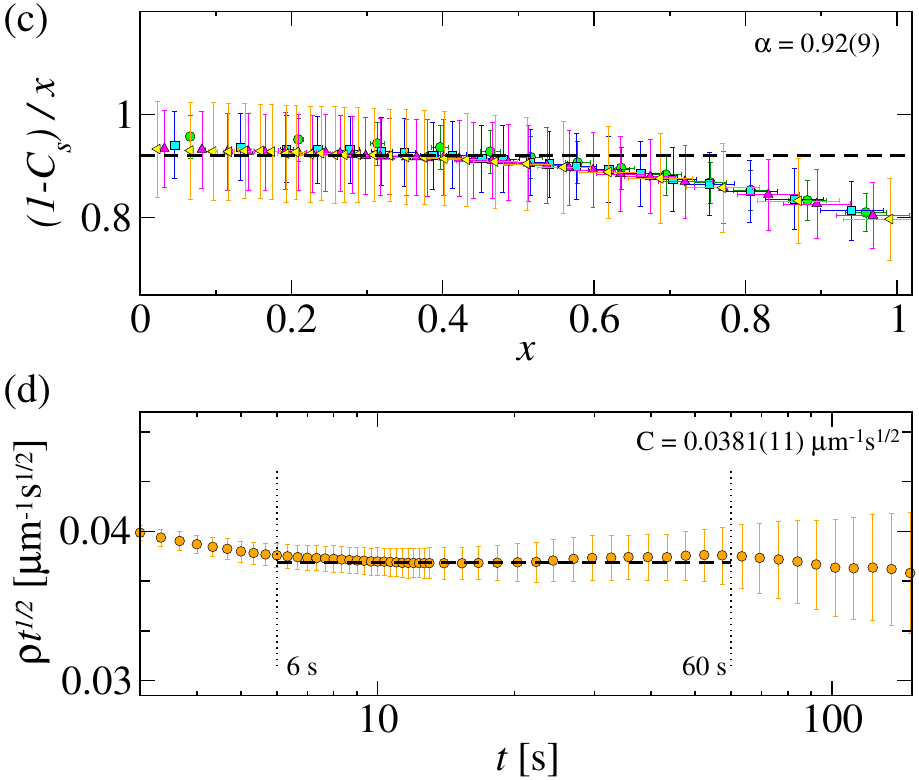} \qquad
        \includegraphics[width=6.5cm]{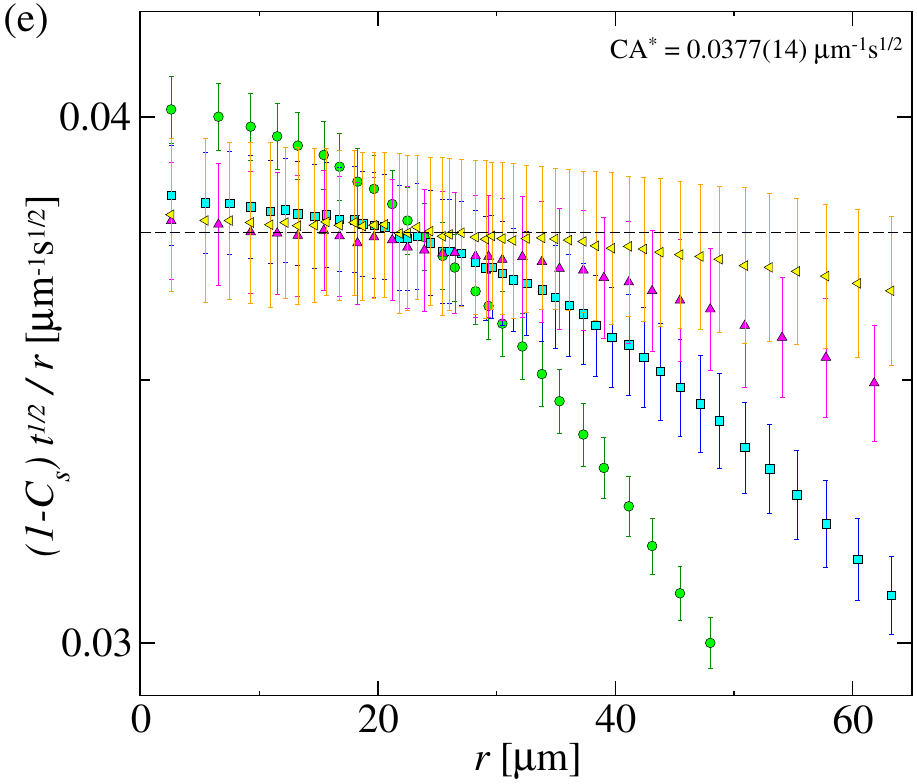} \qquad
        \caption{(a) Spatial two-point correlator $C_s(\bm{r}, t)$ [\eqref{eq-space-covespacial}] versus $r = |\bm{r}|$ at different times. Data were obtained by rotational average (averaging values of $C_s(\bm{r}, t)$ at different $\bm{r}$ with the same $r=|\bm{r}|$), data binning, and ensemble average. Error bars indicate one standard deviation. Dashed line indicates $C_s(\bm{r},t) = 0.2$. Inset shows the growth of $l(t)$, where dashed red line shows $l(t) = Bt^{1/2}$ with $B=24\unit{\mu{}m~s^{-1/2}}$ estimated below \eqref{eq-space-universalamplitude}. (b) Dynamic scaling hypothesis is confirmed from the time-independence of $C_s(\bm{r}, t)$ versus $x = r/l(t)$. Grey dashed line represents the Porod law, $C_s(\bm{r}, t) \simeq 1 - \alpha x$, with $\alpha = 0.92(9)$ estimated from (c). Inset of (b) is a zoom in on the region $x \ll 1$ for the main plot. (c) Estimation of $\alpha$ by reading $(1-C_s)/x$ for $x \rightarrow 0$; dashed line indicates the mean value, $0.92$, of the ordinates in the range $x \leq 0.5$. (d) Plateau of the rescaled interface density $\rho t^{1/2}$ for $6\unit{s} \lesssim t \lesssim 60\unit{s}$; dashed line indicates the mean value $0.0381\unit{\mu m^{-1}s^{1/2}}$. (e) Plot of $(1-C_s)t^{1/2}/r$ versus $r$ from which the amplitude $CA^*$ is estimated in the limit $r \rightarrow 0$. Dashed line indicates the average value $CA^* = 0.0377\unit{\mu m^{-1}s^{1/2}}$. Legends in (b), (c), and (e) are the same as that in (a).}
        \label{Fig_Space-covariance}
        \end{figure*}

    \subsection{Growth law and the Bray-Humayun amplitude}
    \label{Sec_Space_subsec_covariance}

%     In this section, we study the spatial correlator
%     
%         \begin{equation}
%         C_s(\bm{r}, t) = \langle \langle \phi(\bm{r}', t)\phi(\bm{r}' + \bm{r}, t) \rangle_{\bm{r}'}\rangle_\mathrm{e},
%         \label{eq-space-covespacial}
%         \end{equation}
%     where $\langle ... \rangle_{\bm{r}'}$ denotes the spatial average over $\bm{r}'$ and $\langle ... \rangle_\mathrm{e}$ the ensemble average. Figure~\ref{Fig_Space-covariance}(a) shows $C_s(\bm{r}, t)$ against $r = |\bm{r}|$ at different $t$. The larger $t$ is, the more slowly $C_s(\bm{r}, t)$ decays, because of the growth of the typical domain size $l(t)$. Here we set it to be the length at which $C_s(\bm{r}, t) = 0.2$ with $|\bm{r}| = l(t)$. The inset of \figref{Fig_Space-covariance}(a) displays how it grows with $t$, showing a clear power law
    
    The two-point spatial correlator, 
    
         \begin{equation}
         C_s(\bm{r}, t) = \langle \langle \phi(\bm{r}', t)\phi(\bm{r}' + \bm{r}, t) \rangle_{\bm{r}'}\rangle_\mathrm{e},
         \label{eq-space-covespacial}
         \end{equation}
    where $\langle ... \rangle_{\bm{r}'}$ denotes the spatial average and $\langle ... \rangle_\mathrm{e}$ the ensemble average, is shown as function of $r = |\bm{r}|$ in Fig.~\ref{Fig_Space-covariance}(a). The larger the $t$, the slower the $C_s$ decay because of the domain growing size $l(t)$: here set it as the length at which $C_s(\bm{r}, t) = 0.2$ for $|\bm{r}| = l(t)$. The inset of \figref{Fig_Space-covariance}(a) displays the growth law with $z = 2$,

        \begin{equation}
        l(t) \simeq Bt^{1/2} \qquad (\xi \ll l(t) \ll L_{\textrm{sys}}),
        \label{eq-space-l}
        \end{equation}
    with a nonuniversal amplitude $B$ discussed below.%, and the dynamic exponent $z = 2$. %Fitting \eqref{eq-space-l} to the data for $6\unit{s} \lesssim t \lesssim 60\unit{s}$, in consistency with the fit for $\rho(t)$ [\eqref{eq-space-1z}], gives $1/z = 0.503(23)$ or $z = 1.99(9)$. %Now we test the dynamic scaling hypothesis. If it holds, one can define a time-independent scaling function $\mathscr{F}(\cdot)$ such that:
    
    A time-independent scaling function $\mathscr{F}(\cdot)$,
        \begin{equation}
        C_s(\bm{r}, t) \simeq \mathscr{F}(x) \quad (x = |\bm{r}|/l(t)),
        \label{eq-space-scalingfunctionF}
        \end{equation}
    is defined when dynamic scaling holds, as it does in the experiment within the accuracy of $\delta C_s(x,t) \approx 0.07$ (standard deviation) for $t \lesssim 21\unit{s}$ and $x \leq 3$ [see Fig.~\ref{Fig_Space-covariance}(b)]. Moreover, the Porod's law \cite{Porod51_combo},
    
    %The functional form of $\mathscr{F}(\cdot)$ is universal in Model A \cite{Bray02} and depends on $d$, $n$, and possibly on the presence of long-range correlations in initial conditions \cite{Bray91_correlatedIC}. Figure~\ref{Fig_Space-covariance}(b) shows that the scaling hypothesis indeed holds in our TNLC data, within the accuracy of $\delta C_s(x,t) \approx 0.07$ (standard deviation) for $t \lesssim 21\unit{s}$ and $x \leq 3$. 

%   Moreover, the linear decay:
%          
%       \begin{equation}
%       C_s(\bm{r}, t) \simeq 1 - \alpha x \qquad (x = |\bm{r}|/l(t) \ll 1),
%       \label{eq-space-covshort}
%       \end{equation}
%   known as the Porod law \cite{Porod51_combo},

        \begin{equation}
        C_s(\bm{r}, t) \simeq 1 - \alpha x \qquad (\textrm{for } x \ll 1),
        \label{eq-space-covshort}
        \end{equation}
   is emphasized in the inset of \figref{Fig_Space-covariance}(b). The angular coefficient $\alpha$, estimated from the convergence of $(1 - C_s)/x$ for $x \rightarrow 0$ at a late time [$t = 21.33\unit{s}$; see \figref{Fig_Space-covariance}(c)], reads:
   %For data at a late time, specifically $t = 21.33\unit{s}$, averaging the values of the ordinates in the plateau region $x \leq 0.5$ reads:
 
        \begin{equation}
        \alpha = 0.92(9). 
        \label{eq-space-alpha}
        \end{equation}
    
    We also remind that Bray and Humayun \cite{Bray_Humayun93_amplitudes} exactly derived $C_s(\bm{r}, t)$ at $r \ll l(t)$, for $n \leq d$: 
  
        \begin{equation}
        C_s(\bm{r}, t) \simeq 1 + A(d,n)\rho_{\textrm{Euc}}(t) r^n \qquad (\xi \ll r \ll l(t)),
        \label{eq-space-covshortbray}
        \end{equation}
        where $\rho_{\textrm{Euc}}$ is the interface density measured with the Euclidian metric; $A(d,n)$ is the universal Bray-Humayun (BH) prefactor \cite{Bray_Humayun93_amplitudes}: 
        
        \begin{equation}
        A(d,n) = \pi^{(n/2) - 1} \frac{ \Gamma(-n/2)\Gamma(d/2)\Gamma^2((n+1)/2)}{\Gamma((n+d)/2)\Gamma(n/2)},
        \label{eq-space-amplitude}
        \end{equation}
        for odd $n$, where $\Gamma(\cdot)$ denotes the gamma function.
        For our system, $d=2$ and $n=1$, so that $A(2,1) = -4/\pi$.
        
    To assess \eqref{eq-space-covshortbray} in the experiment, we convert $\rho_{\textrm{Euc}}$ to $\rho$ in the Manhattan metric because this metric (the Manhattan one) was used in \figref{Fig_Space-interface}. Thus, $\rho_{\textrm{Euc}} = L'_{\textrm{Euc}}/L_x L_y$ is replaced by $\rho = L'_{\textrm{Man}}/L_x L_y$, where $L'_{\textrm{Euc}}$ ( $L'_{\textrm{Man}}$) is the interface length in the Euclidean (Manhattan) ruler. Noting that $L'_{\textrm{Man}}/L'_{\textrm{Euc}} = 4/\pi$ holds for isotropic space \cite{Bray_Humayun93_amplitudes}, \eqref{eq-space-covshortbray} becomes:
 
        \begin{equation}
        C_s(\bm{r}, t) \simeq 1 - A^* \rho(t) r \qquad (\xi \ll r \ll l(t)),
        \label{eq-space-covuniversalmanhatam}
        \end{equation}
    with the BH amplitude $A^* = 1$.

    %Now we test such prediction experimentally. For that, we focus on the nonuniversal coefficient $B$ in \eqref{eq-space-l}, as well as the coefficient $C$ we introduce here to the power law of $\rho(t)$, $\rho(t) \simeq C t^{-1/2}$ (\figref{Fig_Space-interface}). Substituting this and \eqref{eq-space-l} to \eqref{eq-space-covuniversalmanhatam}, we have
    
    We test such prediction as follows. Defining $C$ as the prefactor of $\rho(t) \simeq C t^{-1/2}$ (\figref{Fig_Space-interface}), and recalling $B/l(t) \simeq t^{-1/2}$ [\eqref{eq-space-l}], we insert these in \eqref{eq-space-covuniversalmanhatam} to obtain:
    
    \begin{align}
        C_s(\bm{r}, t) &\simeq 1 - CA^* r t^{-1/2} \notag \\
        &= 1-A^*BC \frac{r}{l(t)} \qquad (\xi \ll r \ll l(t)),
        \label{eq-space-covuniversalmanhatam2}
    \end{align}
    and we identify $\alpha = A^*BC$. %First we evaluate $C$ by plotting $\rho(t)t^{1/2}$ against $t$ in \figref{Fig_Space-covariance}(d) and reading its plateau value. Time average of $\rho(t) t^{1/2}$ over $6\unit{s} \lesssim t \lesssim 60\unit{s}$ provides $C = 0.0381(11) \unit{s^{1/2}\mu m^{-1}}$. Next, based on the relation $(1-C_s)t^{1/2}/r \simeq CA^*$ for large $t$ and small $r$, we plot $(1-C_s)t^{1/2}/r$ in \figref{Fig_Space-covariance}(e), read the plateau value in the region $0 \leq r \leq 0.5l(t) \approx 57 \unit{\mu m}$ for $t=21.33\unit{s}$ -- which was also used in the estimation of $\alpha$ in \eqref{eq-space-alpha} -- and obtain $CA^* = 0.0377(14) \unit{\mu m^{-1}}$. Using those two estimates, we obtain
        
      \begin{figure*}[!t]
        \centering
        \includegraphics[width=6.5cm]{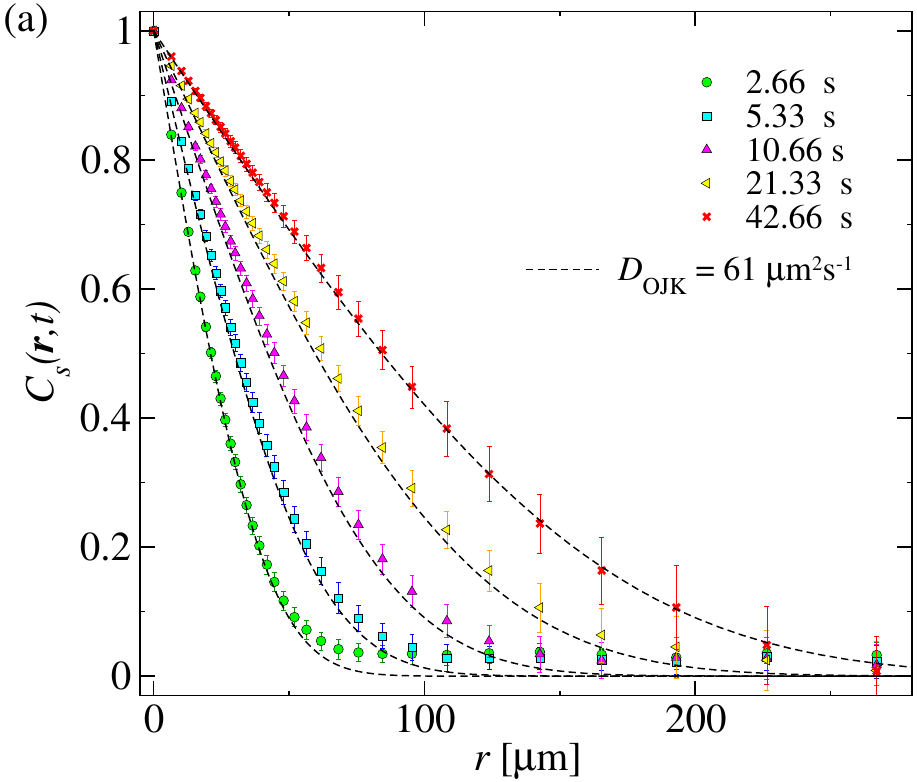} \qquad
        \includegraphics[width=6.5cm]{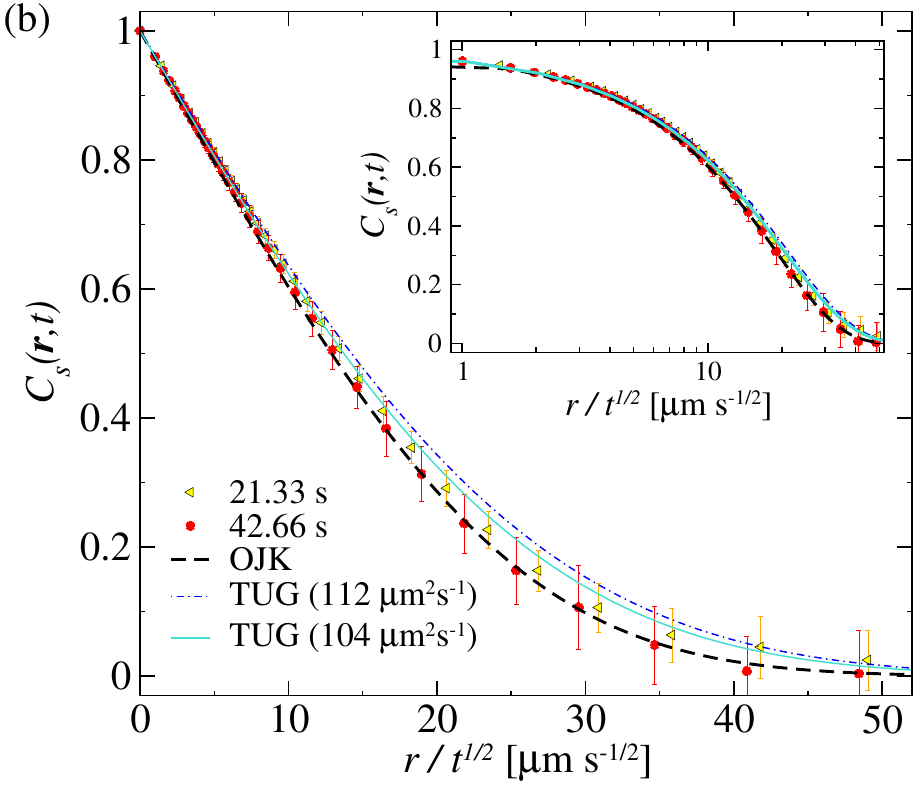} \qquad
        \caption{Two-point spatial correlator $C_s(\bm{r},t)$, \eqref{eq-space-covespacial}, in the experiment (symbols) compared with the forms predicted by the Ohta-Jasnow-Kawasaki (OJK) theory and the theory of unstable growth (TUG) own to Mazenko. The twisted nematic liquid crystal (TNLC) data for $t \leq 21.33\unit{s}$ are identical to those in \figref{Fig_Space-covariance}(a)(b), but the number of data points is reduced for the sake of visibility. Error bars indicate one standard deviation. (a) $C_s(\bm{r},t)$ for the experiment (symbols) at several times and the OJK theoretical form (dashed lines), \eqref{eq-space-OJK-cov}, with $D_{\textrm{OJK}} = 61 \unit{\mu m^2 s^{-1}}$ obtained from the relation with the Allen-Cahn constant: $D_{\textrm{OJK}} = D_{\textrm{AC}}/2$. (b) Rescaled $C_s(\bm{r},t)$ in the experiment compared with both OJK (dashed black line) and TUG (dash-dotted blue and solid turquoise lines) functions. The two TUG curves were obtained with different values of $D_{\textrm{TUG}}$ as indicated in the legend.
        The inset shows the same data in the main plot, but the horizontal axis is displayed in logarithmic scale.
        }
        \label{Fig_Space-covariance-OJKTUG}
        \end{figure*}

    From the plateau in the plot of $\rho(t)t^{1/2}$ versus $t$ [\figref{Fig_Space-covariance}(d)], observed for $6\unit{s} \lesssim t \lesssim 60\unit{s}$~, we read $C = 0.0381(11) \unit{s^{1/2}\mu m^{-1}}$. Similarly from $(1-C_s)t^{1/2}/r$ versus $r$ in \figref{Fig_Space-covariance}(e), the plateau interval $0 \leq r \leq 0.5l(t) \approx 57 \unit{\mu m}$ for $t = 21.33\unit{s}$ -- which was used in the measurement of $\alpha$ in \eqref{eq-space-alpha} -- gives $CA^* = 0.0377(14) \unit{\mu m^{-1}}$, so that we find:

        \begin{equation}
        A^* = 0.99(7)
        \label{eq-space-universalamplitude}
        \end{equation}
    in striking accordance with the BH prediction $A^* = 1$ \cite{Bray_Humayun93_amplitudes}.
    By passing, from $\alpha = A^*BC = 0.92(9)$ in \eqref{eq-space-alpha} and $CA^* = 0.0377(14) \unit{\mu m^{-1}}$ above, one has $B = 24(3) \unit{\mu m s^{-1/2}}$: a value that describes very well the experimental growth law [\figref{Fig_Space-covariance}(a), inset].

    \subsection{Experimental test of Gaussian theories}
    \label{Sec_Space_subsec_Gaussian}

    In scalar 2d Model A, the order parameter field tends to change discontinuously in the scaling limit when interfaces are crossed. Because such discontinuity is hard to deal with analytically, both OJK and TUG theories replace the order parameter by an auxiliary field that smoothly varies in the whole space.
    OJK does so by describing interfaces as a collection of positions satisfying $u(\bm{r},t)=0$, where $u(\bm{r},t)$ is an auxiliary field such that $\phi(\bm{r},t) = \sgn (u(\bm{r},t))$ \cite{OJK82,Bray02}. Starting from the Allen-Cahn interface motion \cite{AllenCahn79} and relying upon approximating $\nabla_{j}u\nabla_{k}u/|\nabla u|^2 \approx \delta_{jk}/d$ with $j, k = 1, 2, ..., d$, OJK reaches at the diffusion equation for $u(\bm{r},t)$. Assuming $u(\bm{r},0)$ as a Gaussian random field\footnote{
    Note that OJK's assumption of a Gaussian field was recently shown to be a reasonable approximation in $d=2$ by numerical simulations of the TDGL equation \cite{Yeung18}.},
    the form of $\mathscr{F}(\cdot)$ in \eqref{eq-space-scalingfunctionF} \cite{OJK82,Bray02} is\footnote{
    It was shown that the nonlinear terms neglected in the OJK theory by the approximation $\nabla_{j}u\nabla_{k}u/|\nabla u|^2 \approx \delta_{jk}/d$ contribute to a correction to \eqref{eq-space-OJK-cov} in the order of $\mathcal{O}(1/d^2)$ for large $d$ \cite{Emmot98}.}:
 
    % \begin{widetext}
%         \begin{equation}
%         \mathscr{F}_{\textrm{OJK}}(r/t^{1/2}) = \frac{2}{\pi}\sin^{-1}\Bigg[ \exp{\Bigg(-\frac{1}{2}\Bigg(\frac{r}{(4D_{\textrm{OJK}}t)^{1/2}}\Bigg)^2\Bigg)} \Bigg],
%         \label{eq-space-OJK-cov}
%         \end{equation}
%     %\end{widetext}
    
        \begin{equation}
        \mathscr{F}_{\textrm{OJK}}(r/t^{1/2}) = \frac{2}{\pi}\sin^{-1}\Bigg[\exp{\Bigg ( \frac{-r^2}{8D_{\textrm{OJK}}t}\Bigg ) } \Bigg],
        \label{eq-space-OJK-cov}
        \end{equation}
       with $D_{\textrm{OJK}} = D_{\textrm{AC}}/2$ for $d=2$ \cite{OJK82}.

    We test \eqref{eq-space-OJK-cov} in the experiment, free from fit, using  $D_{\textrm{OJK}} = D_{\textrm{AC}}/2 = 61(2) \unit{\mu m^{2}s^{-1}}$ after \eqref{eq-space-DacTNLC}. Because $C_s(\bm{r}, t) \to M(t)^2$ for $|\bm{r}| \to \infty$ and $M(t) \neq 0$ in the experiment [\figref{Fig_Space-bubble}(a), inset], but $M=0$ for OJK, comparison is restricted to $C_s(\bm{r}, t) \gtrsim M(t)^2$. Figure \ref{Fig_Space-covariance-OJKTUG}(a) compares the OJK form with TNLC data in the region $C_s(\bm{r}, t) \gtrsim M(t)^2 \approx 0.04$ for $t=42.66\unit{s}$, and $\gtrsim 0.02$ for shorter times. We find noteworthy agreement not only in the initial linear decay (Porod's regime), but also in the interdomain length scale beyond it. Agreement is better for larger times. We also may look at the TNLC growth law, $l(t) \simeq B t^{1/2}$ with $B = 24(3) \unit{\mu m s^{-1/2}}$ (Sec.~\ref{Sec_Space_subsec_covariance}), and compare it with the OJK outcome, $l(t) = \sqrt{-8D_{\textrm{OJK}}\ln[\sin(\pi/10)]}t^{1/2}$ from $\mathscr{F}_{\textrm{OJK}}(l(t)/t^{1/2}) = 0.2$ in \eqref{eq-space-OJK-cov}; this comparison yields $D_{\textrm{OJK}} = 64(18) \unit{\mu m^{2}s^{-1}}$ in compatibility with $D_{\textrm{AC}}/2 = 61(2) \unit{\mu m^{2}s^{-1}}$. 
    
    Now we turn attention to TUG. Developed from the $d$-dimensional TDGL equation, TUG introduces an auxiliary field $m(\bm{r},t)$ interpreted as the shortest distance between $\bm{r}$ and the nearby interface\footnote{More specifically, the relation between the continuous order parameter field $\phi$ and the auxiliary field $m$ is obtained by requiring $\frac{d^2 \phi(m)}{dm^2} = \frac{dV(\phi)}{d\phi}$, where $V(\phi)$ is a symmetric double-well potential, with the boundary conditions $\lim_{m \rightarrow \pm \infty} \phi(m) = \pm 1$ and $\phi(0) = 0$. This is the equilibrium condition for the TDGL equation, with $m$ playing the role of the coordinate normal to the interface \cite{Bray02}.}. Assuming $m(\bm{r},t)$ as a Gaussian field, TUG predicts that $\mathscr{F}(\cdot)$ in \eqref{eq-space-scalingfunctionF} is given by $\mathscr{F}(r/t^{1/2}) = \mathscr{F}_\textrm{TUG}(g)$, which satisfies the following differential equation \cite{Mazenko89,Mazenko90}:

    \begin{widetext}
         \begin{equation}
        \frac{d^2\mathscr{F}_{\textrm{TUG}}(g)}{dg^2} + \Bigg( \frac{d-1}{g} + g\mu(d) \Bigg)\frac{d\mathscr{F}_{\textrm{TUG}}(g)}{dg} + \tan\Bigg(\frac{\pi}{2}\mathscr{F}_{\textrm{TUG}}(g)\Bigg) = 0.
        \label{eq-space-TUG-scalingfunctionF} 
        \end{equation}
    \end{widetext}
    with $\mathscr{F}_{\textrm{TUG}}(0) = 1$, $g = r/(4D_\textrm{TUG}t)^{1/2}$, and $D_\textrm{TUG}$ is a constant. The value $\mu(d)$ is $\mu(2) \approx 1.104$ \cite{Mazenko89,Mazenko90} for the case of interest here\footnote{Equation~(\ref{eq-space-TUG-scalingfunctionF}) admits two linearly independent solutions for $g \gg 1$, one with a Gaussian decay and the other with an algebraic one. The general solution is a linear combination of both: $\mathscr{F}_{\textrm{TUG}}(g \gg 1) = F_0g^{\big(d-\frac{\pi}{2\mu}\big)}\exp{\Big(\frac{-\mu}{2}g^2\Big)} + F_1g^{-\frac{\pi}{2\mu}}$, where $F_0$ and $F_1$ are constants \cite{Mazenko90}. The power-law solution in TUG should be neglected to guarantee the integrability of the associated power-spectrum. The value of $\mu(d)$ in TUG is then given in such a way that $\mathscr{F}_{\textrm{TUG}}(g \gg 1)$ yields a solution with $F_1 = 0$.}.
 
    To test TUG form, we numerically integrated \eqref{eq-space-TUG-scalingfunctionF} by a Euler-Heuns discretization with a uniform step $\delta g = 10^{-6}$, in $d = 2$, and boundary conditions $\mathscr{F}_{\textrm{TUG}}(0) = 1$, $\frac{d\mathscr{F}_{\textrm{TUG}}(g)}{dg}|_{(g = 0)} = -(2/\pi)^{1/2}$. After, to determine $D_\textrm{TUG}$, we rely on the Porod's regime $g \ll 1$ of the TUG correlator \cite{Mazenko90,Mazenko91_2}:
        \begin{equation}
        \mathscr{F}_{\textrm{TUG}}(g) \simeq 1- \Bigg(\frac{1}{D_{\textrm{TUG}}2\pi(d-1)}\Bigg)^{1/2}\frac{r}{t^{1/2}}.
        \label{eq-space-TUG-scalingfunctionFshort}
        \end{equation}
    Relating \eqref{eq-space-TUG-scalingfunctionFshort} with $C_s(\bm{r},t) \simeq 1 - (\alpha/B)(r/t^{1/2})$ [from Eqs.~\pref{eq-space-l} and \pref{eq-space-covshort}], we find $D_{\textrm{TUG}} = (B/\alpha\sqrt{2\pi})^2 = 1/2\pi(CA^*)^2 = 112(8)\unit{\mu m^2 s^{-1}}$.

    Figure~\ref{Fig_Space-covariance-OJKTUG}(b) compares the experimental data with $\mathscr{F}_{\textrm{TUG}}(g)$ for $D_{\textrm{TUG}} = 112 \unit{\mu m^2 s^{-1}}$ (the most probable value), as well as with $\mathscr{F}_{\textrm{OJK}}(r/t^{1/2})$ for $D_{\textrm{OJK}} = 61 \unit{\mu m^2 s^{-1}}$. For the sake of visibility, we show only data for $t = 21.33\unit{s}$ and $42.66\unit{s}$ because they are in the core of the $z = 2$ scaling regime (see \figref{Fig_Space-interface}). Figure~\ref{Fig_Space-covariance-OJKTUG}(b) suggests that $\mathscr{F}_{\textrm{OJK}}(r/t^{1/2})$ describes the experiment better than the TUG counterpart; however, by adjusting the value of $D_\textrm{TUG}$ within the uncertainty, specifically by setting $D_{\textrm{TUG}} = 104\unit{\mu m^2 s^{-1}}$, the form $\mathscr{F}_{\textrm{TUG}}(g)$ can also be made reasonably close to the data [\figref{Fig_Space-covariance-OJKTUG}(b), inset]. Thus, both the OJK and TUG theories are good descriptors (within the statistical accuracy) for the TNLC correlator  and this includes the description of length scales larger than those in the Porod's regime. Unlike $D_\mathrm{OJK}$, $D_\mathrm{TUG}$ does not seem to have a trivial connection with $D_\mathrm{AC}$, albeit it can also be reliably evaluated by fitting the experimental correlator  to the TUG Porod's regime, \eqref{eq-space-TUG-scalingfunctionFshort}, as we have done. It is overall noteworthy that OJK performs remarkably well without such fitting.

 %--------------------------------------------------------------
 \section{Time statistics}
 \label{Sec_Time}

   \subsection{Two-time autocorrelation}
    \label{Sec_Time_subsec_ageing}
 
    %In this section, we study the two-time autocorrelator defined by
     
    The two-time autocorrelator,
     
        \begin{equation}
        C_t(t,t_0) = \langle \langle \phi(\bm{r}, t)\phi(\bm{r}, t_0) \rangle_{\bm{r}} \rangle_\mathrm{e} \quad (t \geq t_0),
        \label{eq-time-cov}
        \end{equation}
    is anticipated \cite{Furukawa89} to decay in the coarsening stages (for most of the cases\footnote{An exception is the 1dXY model whose $C_t(t/t_0)$ decay is exponential \cite{Newman90_soc}.}) as \cite{Bray02}:
    
        \begin{equation}
        C_t(t,t_0) \simeq \mathscr{H}(t/t_0), \quad \mathscr{H}(y) \sim y^{-\lambda/z} \quad (y \gg 1),
        \label{eq-time-covscaling}
        \end{equation}
        for large $t_0$ and $t$, where $\mathscr{H}(y)$ is a universal scaling function and $\lambda$ is the autocorrelation exponent introduced by Fisher and Huse (FH) \cite{Fisher88}. As described in Sec.~\ref{Sec_Intro}, 2d Model A theories suggest: $\lambda_{\textrm{FH}} = 5/4$ \cite{Fisher88}; $\lambda_{\textrm{OJK}} = 1$ \cite{Yeung90}; $\lambda_{\textrm{TUG}} \approx 1.2887$ \cite{Liu91}. 
        
    The scaling function $\mathscr{H}(y)$ in the OJK model is \cite{Yeung90,Bray02}:
            \begin{equation}
            \mathscr{H}_{\textrm{OJK}}(y) = \frac{2}{\pi}\sin^{-1}\Bigg[\Bigg(\frac{4y}{(1+y)^2}\Bigg)^{d/4} \Bigg].
            \label{eq-time-Hojk}
            \end{equation}
    
    For TUG, the function $\mathscr{H}_{\textrm{TUG}}(\bm{g}, \tau(y))$, analogously defined from the two-point spatiotemporal correlator $\langle \langle \phi(\bm{r'}+\bm{r}, t)\phi(\bm{r'}, t_0) \rangle_{\bm{r'}} \rangle_\mathrm{e}$, with $\bm{g} = \bm{r}/(4D_\textrm{TUG}t)^{1/2}$ and $\tau(y) = (\ln y)/4$, satisfies the differential equation \cite{Liu91}:
    \begin{widetext}
            \begin{equation}
            \frac{\partial \mathscr{H}_{\textrm{TUG}}(\bm{g}, \tau(y))}{\partial \tau} =             
            \frac{1}{\mu}\nabla_{\bm{g}}^2 \mathscr{H}_{\textrm{TUG}} + 
            2\bm{g} \cdot \nabla_{\bm{g}} \mathscr{H}_{\textrm{TUG}} + 
            \frac{1}{\mu}\tan \bigg(\frac{\pi}{2}\mathscr{H}_{\textrm{TUG}}\bigg).
            \label{eq-time-Htug}
            \end{equation}
%---- Function written in the isotropy case
%             \begin{equation}
%             \frac{\partial \mathscr{H}_{\textrm{TUG}}(g, \tau(y))}{\partial \tau} = \Bigg[\frac{\partial^2 \mathscr{H}_{\textrm{TUG}}}{\partial g^2} + \bigg(\frac{d-1}{g} + \mu g\bigg)\frac{\partial \mathscr{H}_{\textrm{TUG}}}{\partial g} + \tan \bigg(\frac{\pi}{2}\mathscr{H}_{\textrm{TUG}}\bigg)\Bigg]\frac{1}{\mu} + g\frac{\partial \mathscr{H}_{\textrm{TUG}}}{\partial g},
%             \label{eq-time-Htug}
%             \end{equation}
%             with $g = r/(4D_\textrm{TUG}t)^{1/2}$, $\tau(y) = \ln(y)/4$, and whose initial condition \cite{Liu91} is $\mathscr{H}_{\textrm{TUG}}(g,0) = \mathscr{F}_{\textrm{TUG}}(g)$ given in \eqref{eq-space-TUG-scalingfunctionF}. We focus interest on $\mathscr{H}_{\textrm{TUG}}(0, \tau(y))$.
    This equation can be integrated numerically 
%            Integration is performed by using finite differences (CHECK)
    from the initial condition $\mathscr{H}_{\textrm{TUG}}(\textbf{g},0)$, which is just the vectorial analogue of $\mathscr{F}_{\textrm{TUG}}(g)$ that we also numerically obtained in \eqref{eq-space-TUG-scalingfunctionF}. Here, we are interested in $\mathscr{H}_{\textrm{TUG}}(y) = \mathscr{H}_{\textrm{TUG}}(0, \tau(y))$.

    % --- Fig. Covariancia temporal
    \begin{figure*}[!ht]
        \centering
        \includegraphics[width=6.5 cm]{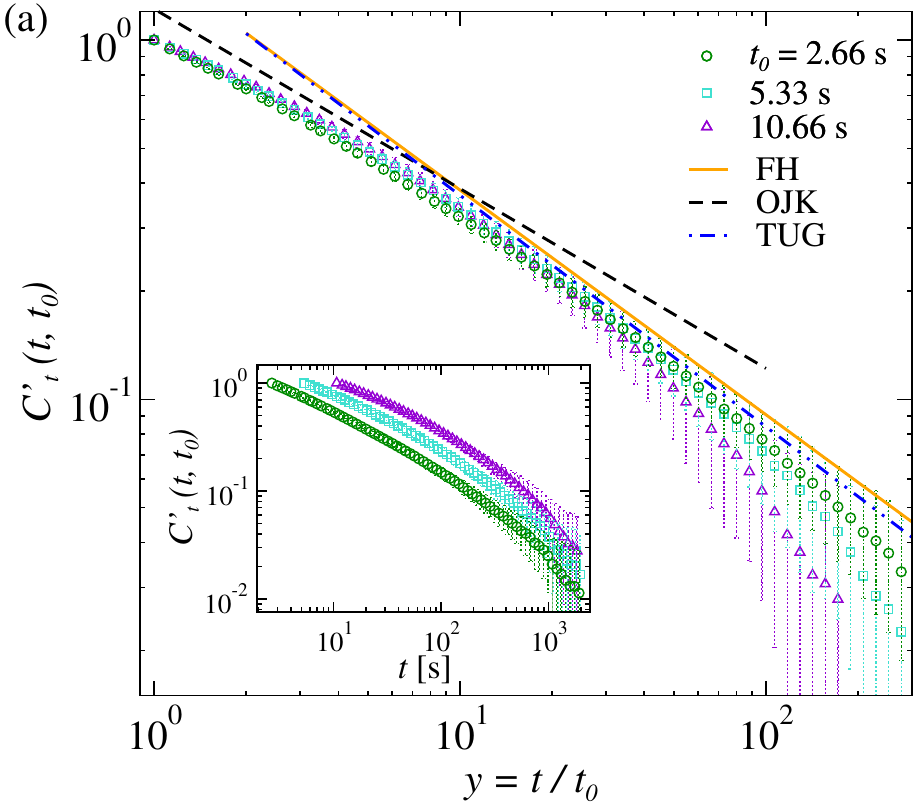}  \qquad
        \includegraphics[width=6.5 cm]{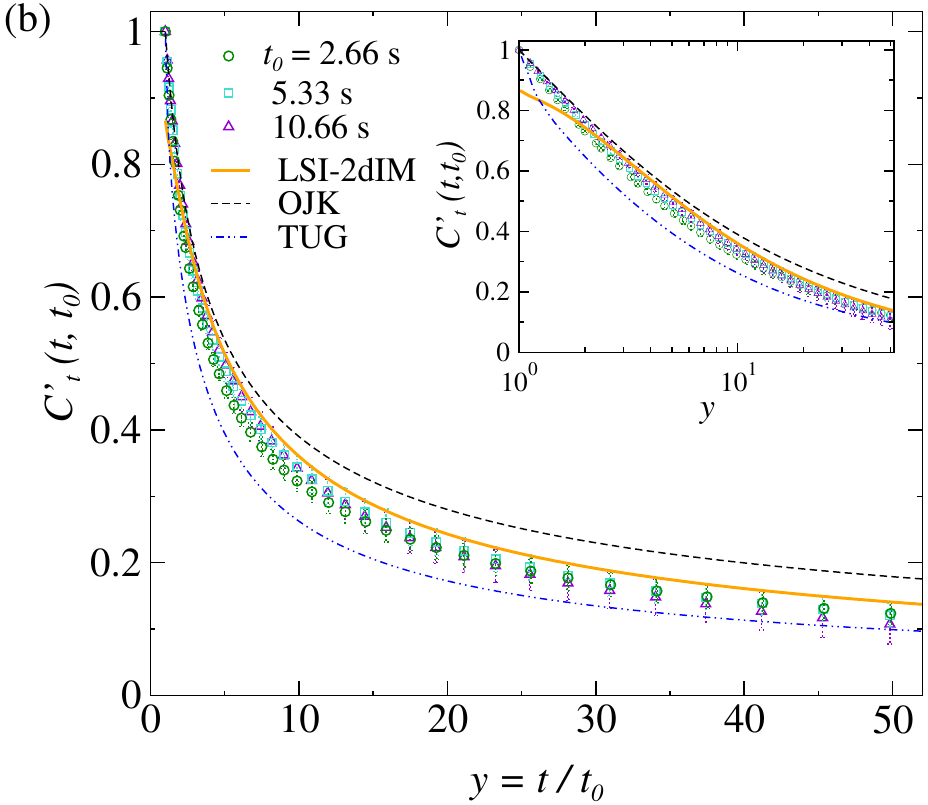} \qquad
        \caption{(a) Two-time autocorrelators $C'_t(t,t_0)$ [\eqref{eq-time-cov-prime}] for the twisted nematic liquid crystal (TNLC) experiment (symbols) versus $t$ (inset), and $y=t/t_0$ (main panel) according to the dynamic scaling hypothesis. The solid orange, dashed black, and dash-dotted blue lines are guides to the eyes whose slopes, $-\lambda/z$, are set by the Fisher-Huse (FH; $-5/8$), Ohta-Jasnow-Kawasaki (OJK; $-1/2$), and the theory of unstable growth (TUG; $\approx -0.644$) predictions, respectively. (b) Same TNLC data compared with forms for the following theoretical predictions: $\mathscr{H}_{\textrm{LSI}}(y)$ [solid orange line; \eqref{eq-time-Hlis} with $E_1 = -0.601$, $E_2 = 3.94$, $E_3 = 0.517$, and $\lambda=\lambda_\textrm{FH}$, which is a solution of the local scaling invariance (LSI) theory that describes the two-dimensional Ising model (2dIM) correlator (see text)], $\mathscr{H}_{\textrm{OJK}}(y)$ [dashed black line; \eqref{eq-time-Hojk} derived from the OJK theory], and $\mathscr{H}_{\textrm{TUG}}(y)$ [dashed-dotted blue line; \eqref{eq-time-Htug} derived from the TUG theory]. Inset displays the same data, but the horizontal axis is displayed in logarithmic scale. In both the panels, error bars indicate one standard deviation.}
        \label{Fig_Time-covariance}
    \end{figure*}

    The LSI theory is based on the hypothesis that correlators transform covariantly under the conformal group (that is, under space-time scale transformations set by the dynamic exponent $z$) \cite{Henkel02}. For 2d scalar Model A, LSI theory yields \cite{Picone04}:
    
    %Finally, the LSI theory, built on the hypothesis that correlators transform covariantly under local scale transformations prescribed by the dynamic exponent $z$ \cite{Henkel02}, predicts the following parametric form for 2d scalar Model A dynamics \cite{Henkel04}:
            \begin{equation}
            \mathscr{H}_{\textrm{LSI}}(y) = y^{\lambda/2}(y-1)^{-\lambda}\Phi\Bigg(\frac{y+1}{y-1}\Bigg),
            \end{equation}
    with
                \begin{equation}
                \begin{aligned}
                %\Phi(q)={} & E_2\Big[\Big( \Gamma(d/2)\textrm{ } _2F_1(\lambda - d/2, d/2; 1 + 2\tilde{a} + d/2 -\lambda; -1/q) - \gamma(d/2,E_3q) \Big)q^{-d/2}\Big] + \\            
                %& + E_1\Big[\Gamma(\lambda - 2\tilde{a})\textrm{ }_2F_1(2\lambda - d - 2\tilde{a}, \lambda - 2\tilde{a}; \lambda + 1 -2\tilde{a} -d/2; -1/q)q^{2\tilde{a}-\lambda} - \\  
                %& - \gamma(\lambda -2\tilde{a}, E_3q)q^{2\tilde{a}-\lambda}\Big] + E_1E_3^{-2\tilde{a}}q^{-\lambda}\gamma(\lambda,E_3q) + E_2E_3^{d/2-\lambda}q^{-\lambda}\gamma(\lambda, E_3q) + \\
                %& + E_1E_3^{1-2\tilde{a}}\frac{2\lambda - d - 2\tilde{a}}{\lambda + 1 - 2\tilde{a} -d/2}q^{-\lambda}\Big[(E_3q)^{2\tilde{a}-1}\gamma(\lambda +1 -2\tilde{a}, E_3q) - \gamma(\lambda, E_3q)\Big],
                \Phi(q) = &E_1 q^{1-\lambda} \left[\left( 1+\frac{1}{q} \right)^{3-2\lambda} \Gamma(\lambda-1) - \gamma(\lambda-1, E_3 q) + \frac{1}{E_3 q}\gamma(\lambda, E_3 q) \right] \\
                & \quad +E_2\left[ E_3^{1-\lambda}q^{-\lambda}\gamma(\lambda, E_3 q) + \frac{e^{-E_3 q} - 1 + {}_2F_1(\lambda-1, 1; 3-\lambda; -1/q)}{q} \right],
                \end{aligned}
                \label{eq-time-Hlis}
                \end{equation}
    \end{widetext}
    where $E_1, E_2, E_3$ are constants, $\Gamma$ is the gamma function, $\gamma$ is the lower incomplete gamma function, and $_2F_1$ is the hypergeometric function. Constants were determined from 2dIM simulations after a quench from the paramagnetic state to the lower-temperature $T = 0$ state, with $\lambda = \lambda_{\textrm{FH}}$; they read \cite{Henkel04}: $E_1 = -0.601$, $E_2 = 3.94$ and $E_3 = 0.517$. Hereafter, $\mathscr{H}_{\textrm{LSI}}(y)$ with this set is referred to as the LSI-2dIM correlator. 
      
    Before comparing experiment with theories, it is crucial to subtract effects from a non-vanishing $M(t)$ in the kinetics, otherwise, the asymptotic decay of the two-time correlator in \eqref{eq-time-cov} becomes dependent on which $\phi = \pm 1$ phase equilibrates the system. %To compare the experimental data with those theoretical predictions, we first need to consider the effect of non-vanishing $M(t)$ on the correlator. %This manifests itself most strongly for $t \gg t_0$, for which $C_t(t,t_0) \simeq M(t)M(t_0)$. To subtract this, we define the modified correlator $C'_t(t,t_0)$ as follows:
    Evaluating the magnetization $\hat{M}(t) = \expct{\phi(\bm{r}, t)}_{\bm{r}}$ and the bare correlator $\hat{C}_t(t,t_0) = \expct{\phi(\bm{r}, t)\phi(\bm{r},t_0)}_{\bm{r}}$ for \textit{each} realization, we define a modified correlator $C'_t(t,t_0)$:
    \begin{equation}
        C'_t(t,t_0) = \Expct{\frac{\hat{C}_t(t,t_0) - \hat{M}(t)\hat{M}(t_0)}{1 - \hat{M}(t)\hat{M}(t_0)}}_\mathrm{e}. \label{eq-time-cov-prime}
    \end{equation}
    This correlator satisfies $C'_t(t_0,t_0)=1$ and $C'_t(t \to \infty,t_0) \to 0$, since $\hat{C}_t(t \to \infty,t_0) \to \hat{M}(t)\hat{M}(t_0)$. 
    %Moreover, again we restrict ourselves to $t \lesssim 60\unit{s}$, for which $|M(t)|$ remains small. 
    Note that $C'_t(t,t_0) \approx C_t(t,t_0)$ as long as $\hat{C}_t(t,t_0) \gg \hat{M}(t)\hat{M}(t_0)$, which is roughly $\approx 0.03$ for $t,t_0 \lesssim 60\unit{s}$ for instance.

    %We measure this modified correlator $C'_t(t,t_0)$ from the TNLC data and show the results in \figref{Fig_Time-covariance}. 
    Figure~\ref{Fig_Time-covariance}(a) displays $C'_t(t,t_0)$ in the experiment for $t_0 = 2.66\unit{s}, 5.33\unit{s}$, and $10.66\unit{s}$ against $t$ (inset) and $y = t/t_0$ (main panel) in the double logarithmic scales. From the collapse of the three data sets shown in the main panel, we confirm the dynamic scaling hypothesis as prescribed in \eqref{eq-time-covscaling}. Those data are then compared with the asymptotic power law decay $C_t(t,t_0) \sim (t/t_0)^{-\lambda/z}$ with $\lambda/z = 1/2$ (OJK), $=5/8$ (FH), and $\approx 0.644$ (TUG). After the initial non-algebraic region, the data at intermediate times $2 \lesssim y \lesssim 10$ tend to decay with a OJK-like value prior to their crossover to a faster, asymptotic decay. Using $t_0 = 2.66\unit{s}$ for which the asymptotic power law is most prominent, and $z=2$ shown in \eqref{eq-space-1z}, we find
    \begin{equation}
        %\lambda/z = \begin{cases} 0.51(3) & (2 < y < 10), \\ 0.64(6) & (15 < y < 75). \end{cases}  \label{eq-lambda}
        \lambda = \begin{cases} 1.03(5) & (2 \leq y \leq 10); \\ 1.28(11) & (15 \leq y \leq 75), \end{cases}  \label{eq-lambda}
        %\quad\text{(effective exponent)} \qquad \lambda = \begin{cases} XXX(X) & (3 < y < 10), \\ XXX(X) & (XXX < y < XXX), \end{cases} \quad\text{(power-law fit)}
    \end{equation}
    whose values for the asymptotic TNLC decay are compatible with both the FH conjecture and TUG theory within uncertainty (uncertainties are evaluated from the effective exponents).

    Figure~\ref{Fig_Time-covariance}(b) compares $C'_t(t,t_0)$ in the experiment with the OJK [\eqref{eq-time-Hojk}], TUG [\eqref{eq-time-Htug}], and LSI-2dIM [\eqref{eq-time-Hlis}] functions. %with $E_1 = -0.601$, $E_2 = 3.94$ and $E_3 = 0.517$] theories. 
    %This shows that, Among the predictions, the LSI-2dIM curve best describes the experimental correlator, except for $y \lesssim 2$ because LSI solutions do not describe such regime \cite{Henkel02,Henkel04}. As opposed to the LSI-2dIM prediction, neither OJK or TUG theory describes the experimentally observed autocorrelator, thus revealing that neither of them perform well to describes the universal autocorrelator form in the Model A class. 
    OJK outcome is well above the TNLC data, a result similarly seen in tests with 2dIM simulations \cite{Henkel04}. The TUG form, which to our knowledge has not been confronted with simulations or experiments, exhibits pronounced deviations from the experiment for $y \lesssim 25$ while it approaches to the collapsed correlator from below. This suggests that neither OJK or TUG theory may perform well to describe the universal autocorrelator form in scalar, 2d Model A dynamics. Interestingly, the experimental correlator is best accounted for by the LSI-2dIM theory, except for $y \lesssim 2$ since LSI solutions cannot describe such regime \cite{Henkel02,Henkel04}. The nice agreement with LSI-2dIM can also be appreciated in the $\log \times$ linear plot shown in the inset of \figref{Fig_Time-covariance}(b).

    In an overall view, individual theories cannot account simultaneously for all aspects (or regimes) of the experimental two-time autocorrelator. Indeed, although both the FH conjecture and TUG theory are good descriptors for the slope of the power-law decay, the autocorrelator functional form itself is not captured by TUG, nor OJK. Instead, this form is overall best accounted for by the LSI-2dIM correlator, except for the $y \lesssim 2$ regime; note that LSI-2dIM incorporates the FH assumption, which is also in harmony with our experimental observations.
    
 %--------------------------------------------------------------
   
 \subsection{Local persistence}
 \label{Sec_Time_subsec_persistence}

\begin{figure}[!t]
                \centering
                \includegraphics[width=6.5cm]{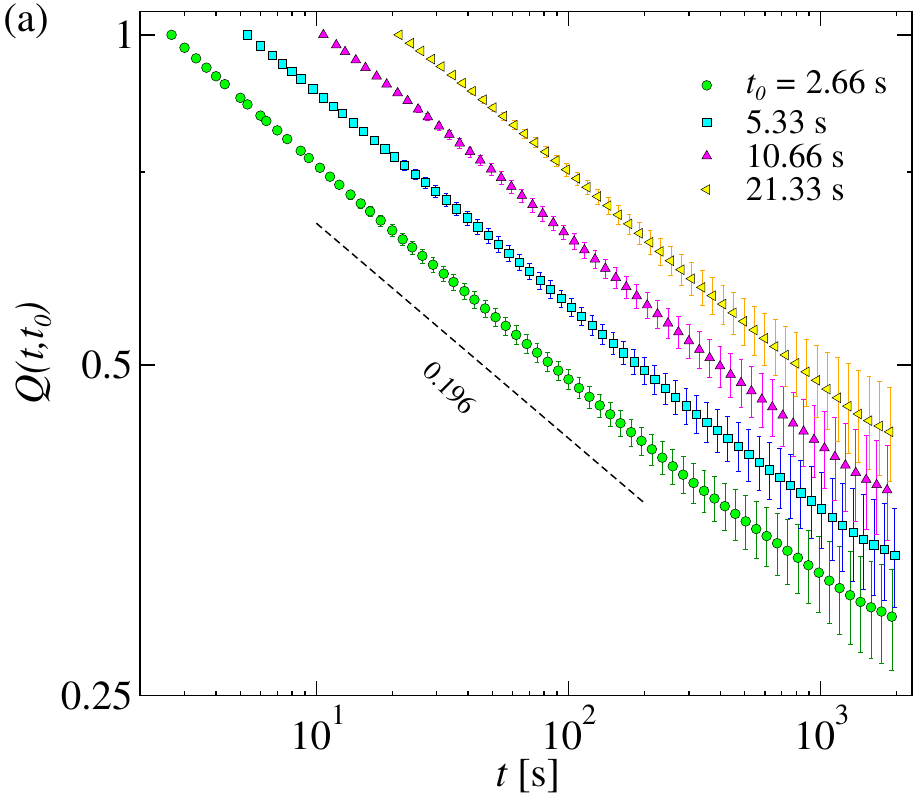}  \qquad
                
                \vspace{0.5cm}
                
                \includegraphics[width=6.5cm]{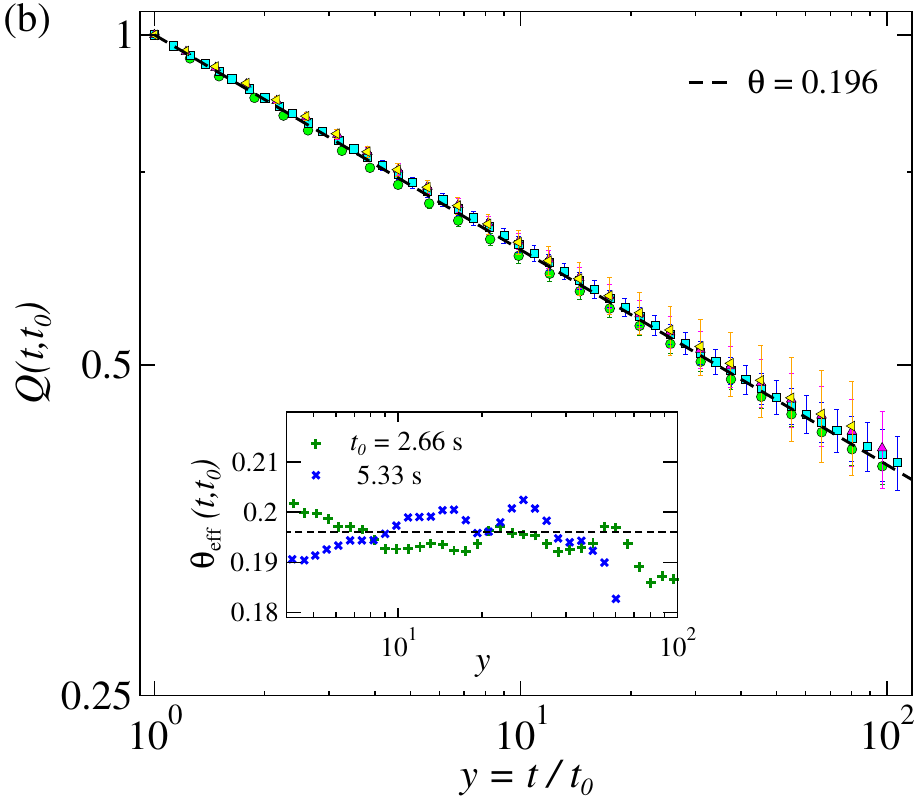} \qquad
                \caption{(a) Local persistence probability $Q(t,t_0)$ for the ordering kinetics of twisted nematic liquid crystals (TNLC), for several reference times $t_0$. Error bars indicate one standard deviation. Dashed line is a guide to eyes whose absolute value of its slope is indicated in the plot. (b) Collapse of $Q$ when plotted versus $y = t/t_0$, and \eqref{eq-time-persistence} with $\theta = 0.196$. Legend of symbols as shown in (a). Inset shows the effective exponent $\theta_{\textrm{eff}}(t,t_0)$ for $t_0 = 2.66\unit{s}$ and $5.33\unit{s}$ data. Dashed line indicates the mean value, 0.196, for the $8 \leq y \leq 50$ interval.}
                \label{Fig_Persistence-probability}
                \end{figure}

        The local persistence probability $Q(t,t_0)$ is the probability that sgn$\big(\phi(\bm{r},\tilde{t})\big)$ does not change during $\tilde{t} \in [t_0, t]$. It typically shows nontrivial algebraic decay \cite{Majumdar99_rev,Bray13_rev},
        
            \begin{equation}
            Q(t,t_0) \sim (t/t_0)^{-\theta} \quad (t \gg t_0),
            \label{eq-time-persistence}
            \end{equation}
            with the persistence exponent $\theta$,
            %that normally does not have direct relationship to standard critical exponents.
            as we can here directly appreciate in the experiment for several values of $t_0$ [\figref{Fig_Persistence-probability}(a)].
            %We measure $Q(t,t_0)$ in the experiment and indeed find the power law \pref{eq-time-persistence} for several values of $t_0$ [\figref{Fig_Persistence-probability}(a)]. 
            The plot $Q(t,t_0)$ versus $y = t/t_0$ yields an excellent data collapse within $\delta Q \approx 0.04$ (standard deviation) for $y \lesssim 10^2$ [\figref{Fig_Persistence-probability}(b)]. 
            The decay is quantified by the exponent, $\theta_{\textrm{eff}}(t,t_0) = - d(\ln Q(t,t_0) )/d(\ln t)$, shown in the inset of \figref{Fig_Persistence-probability}(b) for $t_0 = 2.66\unit{s}$ and $5.33\unit{s}$. Both of the $\theta_{\textrm{eff}}(t,t_0)$ curves fluctuate around their mean values for $8 \leq y \leq 50$. Averaging $\theta_{\textrm{eff}}$ data in this interval gives $0.1938(14)$ and $0.1976(27)$ for $t_0 = 2.66\unit{s}$ and $5.33\unit{s}$, respectively. We also evaluate $\theta$ by applying a least-squares fit individually to $Q(t,t_0)$ data of each realization, constrained to the interval $8 \leq y \leq 50$, and then ensemble averaging; the results are: $\theta = 0.1941(26)$ and $0.199(4)$ for $t_0 = 2.66\unit{s}$ and $5.33\unit{s}$. Integrating these results, we report
            
                % --- Eq.: persistence exponent
                \begin{equation}
                    \theta = \begin{cases}
                    0.196(3) & \mbox{(effective exponent);} \\
                    0.196(4) & \mbox{(power-law fit),}
                    \end{cases}
                \label{eq-time-theta}
                \end{equation}
            where uncertainties include (sum up) both the uncertainty from each estimate and the deviation between the two estimates for $t_0 = 2.66\unit{s}$ and $5.33\unit{s}$. We remark that these results do not significantly change when the interval $8 < y < 60\unit{s}/t_0$ is used, since it corresponds to the interval $t \lesssim 60\unit{s}$ predominantly used in Sec. \ref{Sec_Space}.
            %Results does not significantly change when the interval $8 \leq y \leq 60\unit{s}/t_0$ is used%, which corresponds to the time interval $t \leq 60\unit{s}$ used predominantly in our analysis.
 
                % --- Fig.: Local persistence
  
            Results in \eqref{eq-time-theta} agree with systematic 2dIM outcomes \cite{Blanchard14} ($\theta \approx 0.195$ and $\theta = 0.199(2)$ for free and periodic boundary conditions, respectively), but they significantly differ from $\theta_{\textrm{DIF}} = 3/16 = 0.1875$ \cite{Poplavskyi18} and, hence, from the OJK theory as well \cite{Derrida96_OJK}. This may suggest, based on a concrete experience, the former as the true $\theta$ for 2d Model A systems.
            %As we shall discuss at the end of this section, the consistency of the reported value of $\theta$ will become clear through relation to other characteristic quantities of the 2d Model A dynamics.
            %As we shall discuss at the end of this section, such a result provides unambiguous experimental evidence for $\theta$ in the 2d Model A dynamics.
            
            % --- Fig.: Panels of persistence clusters
            \begin{figure*}[!ht]
            \centering
            \includegraphics[width=0.99\hsize,clip]{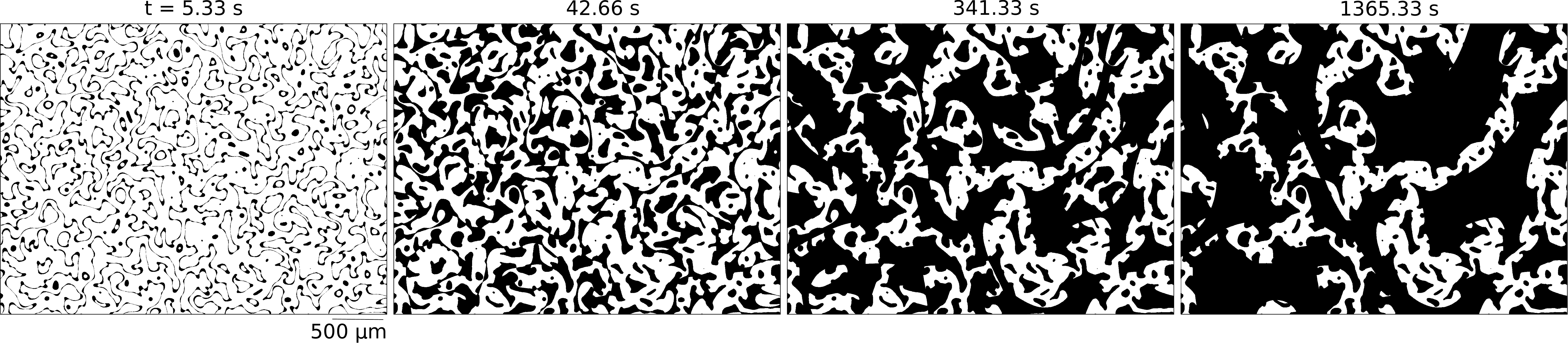}
            \caption{Configurations of the local persistence index $\chi(\bm{r},t)$ (see text) computed in the ordering of twisted nematic liquid crystals (TNLC) for a reference time $t_0 = 2.66 \unit{s}$. The black and white pixels correspond to $\chi = 0$ (nonpersistence) and $1$ (persistence), respectively. See also Supplemental Videos 3 and 4.}
            \label{Fig_Persistence-domains}
            \end{figure*}

            % --- Fig. Fractal dimension and covariance
            \begin{figure*}[!ht]
            \centering
            \includegraphics[width=6.5cm]{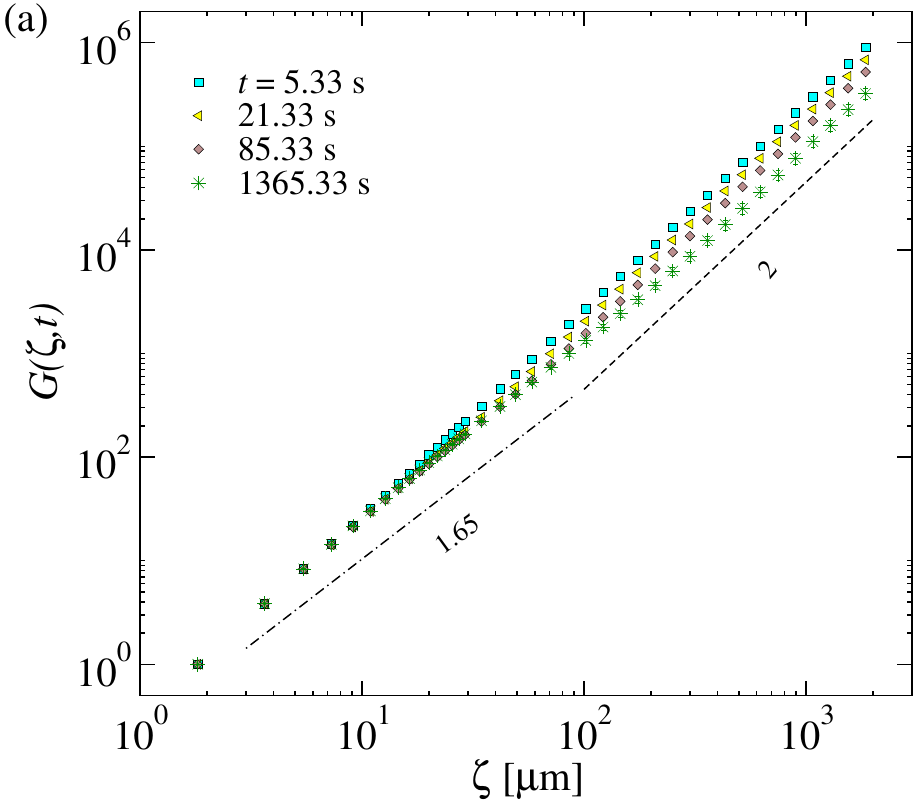}  \qquad
            \includegraphics[width=6.5cm]{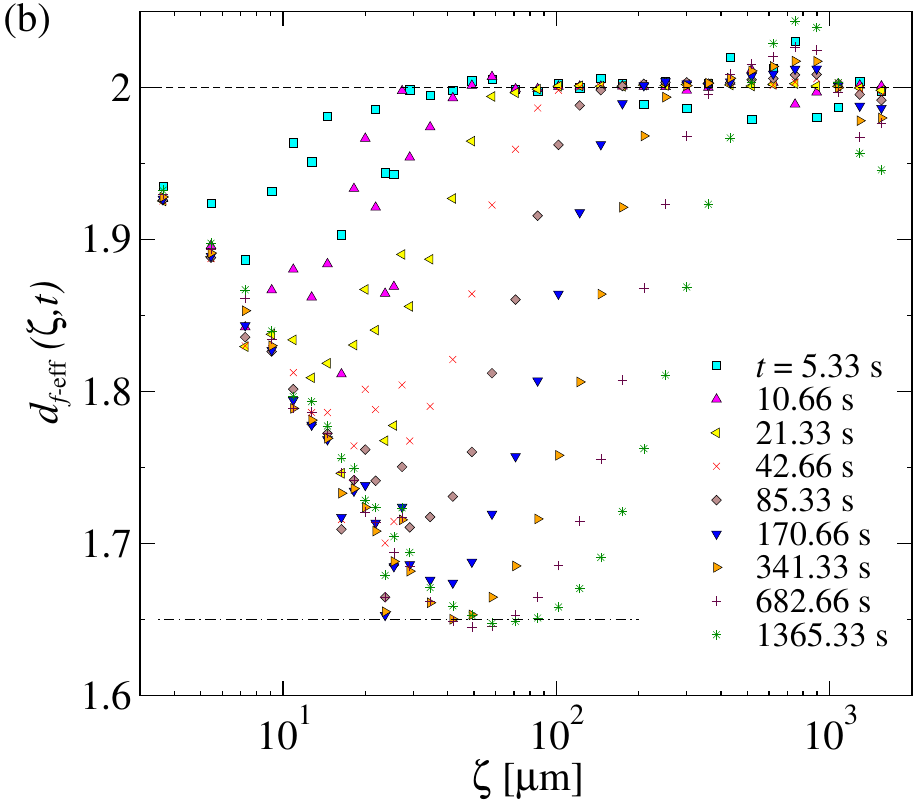} \qquad

                \vspace{0.1cm}

            \includegraphics[width=6.5cm]{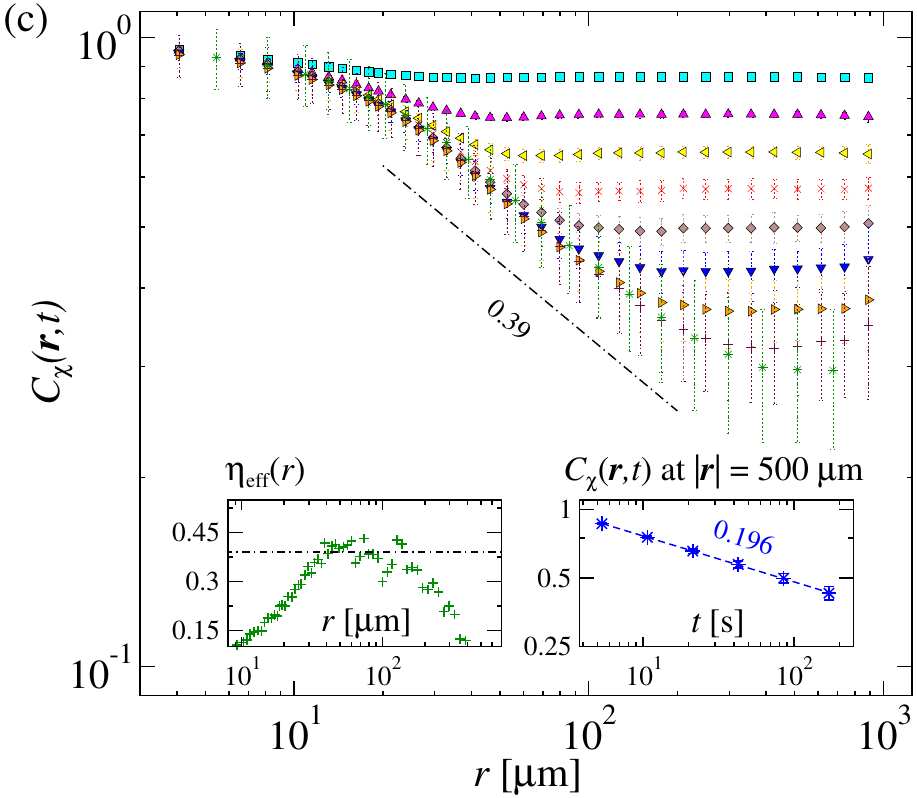} \qquad
            \includegraphics[width=6.5cm]{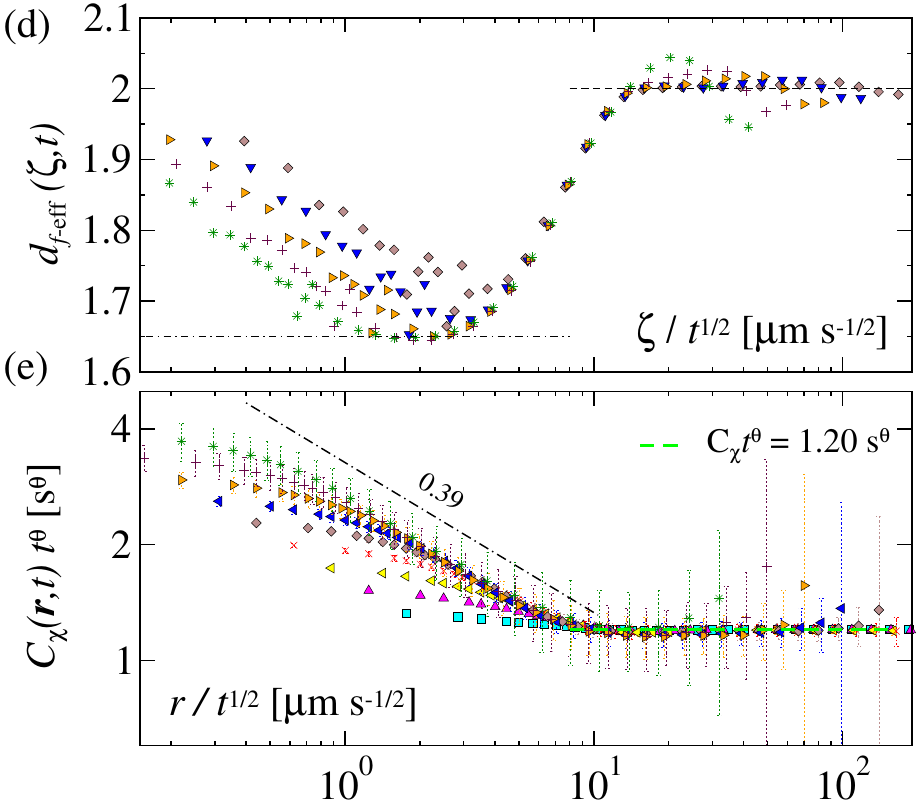} \qquad
            \caption{Statistics of the local persistence in the ordering kinetics of TNLC phases. In all the panels, dash-dotted and dashed lines are guide to eyes; the values of their slopes or ordinates are indicated in the respective plots. (a) Mean counting $G(\zeta,t)$ of $\chi = 1$ (persistence) pixels found inside a square of size $\zeta$ that glides over the configurations of the persistence index $\chi(\bm{r},t)$ (reference time $t_0 = 2.66\unit{s}$). (b) Effective dimension $d_{f\textrm{-eff}}(\zeta,t)$ for the morphology of persistence clusters as a function of $\zeta$. (c) Covariance $C_{\chi}(\bm{r},t)$ [\eqref{eq-time-covpd}] of the index $\chi$ versus $r = |\bm{r}|$. Data were obtained by rotational average (averaging over different $\bm{r}$ with the same $r=|\bm{r}|$), data binning, and ensemble average. The legend of the symbols is given in (b). Left inset depicts the effective exponent $\eta_{\textrm{eff}}(r)$ related to a curve, $t=1365.33\unit{s}$, in the main panel. Right inset shows $C_{\chi}(\bm{r},t)$ in time for a length ($r=500\unit{\mu m}$) much larger than the crossover length $\zeta^*(t)$. (d-e) $d_{f\textrm{-eff}}(\zeta,t)$ and $C_{\chi}(\bm{r},t)$ versus $\zeta/t^{1/2}$ and $r/t^{1/2}$, respectively, following the dynamic scaling hypothesis. Legends in (c),(d),(e) are as that shown in (b). }
            \label{Fig_Persistence-massa}
            \end{figure*}

        % Fractal dimension
        To probe morphological aspects of persistence, we define an index $\chi(\bm{r},t)$ such that $\chi=1$ if sgn$( \phi(\bm{r},\tilde{t}))$ does not change during $\tilde{t} \in [t_0, t]$, with $t_0$ hereafter pinned at $t_0 = 2.66\unit{s}$, and $\chi = 0$ otherwise. The index $\chi(\bm{r},t)$ can only change from $1$ to $0$; such change happens only when an interface crosses the position $\bf{r}$ for the first time since $t_0$.
        %A persistent cluster is a connected path of neighbouring $\chi = 1$ sites.
                 
        Snapshots of $\{\chi(\bm{r},t)\}$ are presented in \figref{Fig_Persistence-domains}; they reveal how a single $\chi=1$ cluster is progressively fragmented because of the interface motion in real space. Surviving clusters develop ragged edges in both internal and external contours resembling fractal objects (see Supplemental Videos 3 and 4). The fractal dimension $d_f$ of this morphology is estimated by combining the gliding-box method \cite{Allain91} and the method in \cite{Manoj00}. Here, a square of side $\zeta$ glides over each $\{\chi(\bm{r},t)\}$ image while counting the number of $\chi = 1$ pixels inside the $i^{th}$ box, $g_i$, with $i = 1, 2, ..., n(r)$; $n(r)$ is the total number of distinct boxes that contain at least one $\chi = 1$ pixel. The mean counting is $G(\zeta,t) = \langle \langle g_i\rangle_{n} \rangle_\mathrm{e}$, where $\langle ... \rangle_{n}$ is the average over $n(r)$ boxes glided on a snapshot at $t$. For fractal objects \cite{Manoj00,Mandelbrot83}: 
            % --- Eq. Mean number counting
            \begin{equation}
            \centering 
            G(\zeta,t) \sim \zeta^{d_f} \quad (a \ll \zeta \ll L_{\textrm{sys}}).
            \label{eq-time-massscale}
            \end{equation}

        In the experiment, $G$ exhibits a crossover from fractal to Euclidian length scales at a certain correlation length $\zeta^*(t)$ [\figref{Fig_Persistence-massa}(a)]. %This $\zeta^*(t)$ can be regarded as a correlation length. 
        The fractal regime is progressively built in time since it is confined to emerge for $a \ll \zeta \ll \zeta^*(t)$, with a growing length, $\zeta^*(t)$, that starts from $\zeta^*(t_0)=0$. This progressive building is captured by $d_{f\textrm{-eff}}(\zeta,t) = d(\ln G(\zeta,t))/d(\ln \zeta)$ [\figref{Fig_Persistence-massa}(b)] which shows a concave dependence with $\zeta$ and forms a plateau around a $t$-dependent minimum. For $\zeta \gg \zeta^*(t)$, $d_{f\textrm{-eff}}$ crossovers to $d = 2$. Finite-size and finite-time effects preclude observation of clear $t$-independent plateau for $\zeta \ll \zeta^*$; however, from the plateau at the minimum of $d_{f\textrm{-eff}}$ for large times, $t=1365\unit{s}$, we roughly read $d_f \approx 1.65(3)$ [\figref{Fig_Persistence-massa}(b)]. The number in the parentheses indicate one standard deviation. %of the results of the power-law fits to the data from the individual realizations.
        %of the effective exponent values within the narrow plateau region. 
        Although $t=1365\unit{s}$ is far from the time interval where $M(t)$ is constant [\figref{Fig_Space-bubble}(a), inset], properties of persistence clusters at long times are reminiscent of those in the early dynamics: this is revealed by the inner structures of persistence domains in the panel at $t=1365\unit{s}$ of \figref{Fig_Persistence-domains}, which are mostly frozen and already present since $t=43\unit{s}$. Note also that the rough estimate $1.65(3)$ is consistent with $d_f = 1.61(5)$ reported for persistence clusters in the 2dIM \cite{Jain00}. 
        
        To evaluate $d_f$ from shorter times, for which $|M(t)|$ is small and $z = 2$ is clearly identified, 
        %for which $|M(t)|$ remains small and most of our results on the Model A behaviour have been identified so far, %with a relationship to $\theta$, 
        we appeal to the correlator $C_{\chi}$,
            \begin{equation}
            C_{\chi}(\bm{r},t) = \frac{\langle\langle \chi(\bm{r'},t) \chi(\bm{r} + \bm{r'},t)\rangle_{\bm{r'}}\rangle_{e}}{\langle\langle \chi(\bm{r'},t) \rangle_{\bm{r'}}\rangle_{e}},
            \label{eq-time-covpd}
            \end{equation}
        which behaves as \cite{Zanette97,Manoj00,Jain00}:
            \begin{equation}
            C_{\chi} \sim \begin{cases}
            r^{-\eta} & (r \ll \zeta^*(t)), \\
            t^{-\theta} & (r \gg \zeta^*(t)),
            \end{cases}            
            \label{eq-time-covpd2}
            \end{equation}
        where $\eta$ is a universal exponent. At the crossover length, $C_{\chi} \sim (\zeta^*)^{-\eta} \sim t^{-\theta}$. Thus, if $\zeta^*(t) \sim t^{1/z}$ holds\footnote{
        It has been shown by Bray and O'Donoghue \cite{Bray00} that, for 1d Potts models, the general law for the characteristic persistence length is $\zeta^* \sim t^{\textrm{max}(1/z, \theta)}$. It turns out that for all integer $q \geq 3$ one has $\theta > 1/z$ \cite{Derrida95,Derrida96}, so that the relation $\zeta^* \sim t^{1/2}$ breaks down.
        For the TNLC experiment, however, $\zeta^* \sim t^{1/2}$ is confirmed in \figref{Fig_Persistence-massa}(d).
        }, then $\eta = z\theta$. Because $G(\zeta,t) \sim \int_\textrm{box} C_{\chi}(\bm{r},t)d^d\bm{r}$ \cite{Manoj00}, where the integration is carried out over a box of side $\zeta$ centered at $\bm{r}=\bm{0}$, integration after \eqref{eq-time-covpd2} yields:
            \begin{equation}
            G(\zeta,t) \sim \begin{cases}
            \zeta^{d -\eta} & (a \ll \zeta \ll \zeta^*(t)); \\
            t^{-\theta}\zeta^d  & (\zeta^*(t) \ll r \ll L_{\textrm{sys}}).
            \end{cases}
            \label{eq-time-masscovpdscale}
            \end{equation}
        Comparing with \eqref{eq-time-massscale}, we identify:
            \begin{equation}
            d_f = d - \theta z,
            \label{eq-time-fractaldimension}
            \end{equation}
        and, consequently, $z\theta \leq d$.

        Before using \eqref{eq-time-fractaldimension} to evaluate $d_f$, we test \eqref{eq-time-covpd2} in the experiment. As shown in \figref{Fig_Persistence-massa}(c), we detect a nearly time-independent decay $C_{\chi}(\bm{r},t) \sim r^{-\eta}$ (arguably from $t= 43\unit{s}$) within a region $20\unit{\mu m} \lesssim r \lesssim \zeta^*(t)$ that extends in time. The exponent $\eta_{\textrm{eff}}(r) = - d(\ln C_{\chi}(r,t))/d(\ln r)$ measured at $t = 1365.33\unit{s}$ provide us with $\eta = 0.39(4)$ after averaging data for $38 \unit{\mu m} \leq r \leq 150\unit{\mu m}$ [see left inset of \figref{Fig_Persistence-massa}(c)]. Note that this value is consistent with $\eta = 0.428(7)$ obtained for persistent 2dIM clusters \cite{Jain00}. We also observe $C_{\chi}(\bm{r},t) \sim t^{-\theta}$ for $r \gg \zeta^*$ in the right inset of \figref{Fig_Persistence-massa}(c), which displays $C_{\chi}(\bm{r},t)$ at $|\bm{r}|=500\unit{\mu{}m}$. Finally, the assumption $\zeta^*(t) \sim t^{1/2}$ is verified by the collapse of $d_{f\textrm{-eff}}(\zeta,t)$ [\figref{Fig_Persistence-massa}(d)] and $C_{\chi}(\bm{r},t)$ [\figref{Fig_Persistence-massa}(e)] when plotted versus $\zeta/t^{1/2}$ and $r/t^{1/2}$, respectively.
        % $r_a \approx 8\unit{\mu m}$ is a microscopic length scale. 
        Deviations from collapsed forms for small lengths are due to effects of microscopic scales, such as $a$ and $\xi$.
        
        Now we safely rely on \eqref{eq-time-fractaldimension} to evaluate $d_f$. Using $\theta$ in \eqref{eq-time-theta} and $z$ in \eqref{eq-space-1z}, we find
        \begin{equation}
            d_f = \begin{cases}
                    1.613(14) & \mbox{(effective exponent);} \\
                    1.609(9) & \mbox{(power-law fit).}
                    \end{cases}
                \label{eq-time-df}    
        \end{equation}
        where the notes in the parentheses indicate the methods used to evaluate $1/z$ and $\theta$ in \eqsref{eq-space-1z} and \pref{eq-time-theta}, respectively. Final estimates are slightly lower than the rough estimate $d_f = 1.65(3)$ (as may be  expected), but, more importantly, they sharply agree with $1.61(5)$ from 2dIM simulations \cite{Jain00}. 
 
        Altogether, experimental results in this Section provide firm evidence that the statistics of the local persistence probability correspond to a universal feature of 2d Model A systems. In addition, one can realize that the presence of a weak phase asymmetry has negligible, or very small, effect on those quantities.
 
 %--------------------------------------------------------------
 \section{Conclusions}
 \label{Sec_Conclusions} 

 % --- Brief recap of the experiment 

  We have studied the nonconserved phase-ordering kinetics of two-dimensional (2d) twisted nematic liquid crystals (TNLC) as a model experimental system to elucidate universal aspects in the Allen-Cahn (Model A) universality class. Exploiting electrically-driven hydrodynamical regimes of liquid crystals, we triggered a genuinely sudden transition between a disordered-like, spatiotemporally chaotic state, called dynamic scattering mode 2 (DSM2), towards a two-phase competing, equilibrium one. The sudden removal of the driving avoids the inexorable slow quench rate of thermally-induced transitions, besides it avoids the issues of metastability, nucleation and nuclei growth, implied in first-order transitions. In addition to its accuracy, the system offers an elegant way of inducing different types of kinetics by transiting between nonequilibrium states while finely tuning thermal effects.

  \begin{table*}[!ht]
        \centering
        \caption{Universal quantities measured in the TNLC experiment}
        \label{table1}
        
        \begin{tabular}{lllll}
        \hline\hline\noalign{\smallskip}
        Quantity            & \multicolumn{2}{l}{Experiment (this work)}             & Theory & References \\
        \noalign{\smallskip}\hline\noalign{\smallskip}
        $1/z$     & $0.505(15), 0.498(6)^{a}$ & \figref{Fig_Space-interface}, \eqref{eq-space-1z}       & $1/2$          & exact predictions \cite{Lifshitz62,AllenCahn79,Bray_Rutenberg94}           \\ \\
        
        $A^*$               & $0.99(7)$ & \eqref{eq-space-universalamplitude} & $1$ & BH prediction \cite{Bray_Humayun93_amplitudes} \\ \\

        $\lambda$ & $1.03(5)$ for short times ($2 \leq y \leq 10$) & \figref{Fig_Time-covariance}(a), \eqref{eq-lambda}  & $1$ & OJK \cite{Yeung90}  \\
         & $1.28(11)$ for late times ($15 \leq y \leq 75$) & & $5/4 = 1.25$ & FH \cite{Fisher88}  \\
         & & & $\approx 1.2887$ & TUG \cite{Liu91}  \\ \\
                               
        $\theta$            & $0.196(3), 0.196(4)^{b}$ & \figref{Fig_Persistence-probability}, \eqref{eq-time-theta}                 &  $\approx 0.195, 0.199(2)^{c}$ & 2dIM simulations \cite{Blanchard14}                                \\ 
        & & & $3/16 = 0.1875$        & DIF-RIC \cite{Poplavskyi18} \& OJK \cite{Derrida96_OJK}                          \\ \\
        
        $d_f$               & $1.65(3), 1.613(14), 1.609(9)^{d}$ & \figref{Fig_Persistence-massa} &  $1.61(5)$  & 2dIM simulations \cite{Jain00}  \\
    \noalign{\smallskip}\hline\hline
    \end{tabular}
    \begin{flushleft} 
%     $^{a}$ In addition to the quantities listed in this table, we also measured the universal scaling function for the spatial correlator (\figref{Fig_Space-covariance-OJKTUG}) and that for the temporal correlator [\figref{Fig_Time-covariance}(b)] and compared with theoretical predictions. \\
     $^{a}$ Both the estimates were obtained from the decay of the interface density, $\rho(t) \sim t^{1/z}$. The time averaged effective exponent and the least-square fit of the power law give $0.505(15)$ and $0.498(6)$, respectively. Please see \eqref{eq-space-1z} and text for detail. \\
     $^{b}$ The estimate $\theta = 0.196(3)$ was obtained from the time average of the effective exponent; $0.196(4)$ from the least-square fit of the power law. See \eqref{eq-time-theta} and text for detail. \\
     $^{c}$ The two values $\theta \approx 0.195$ and $0.199(2)$, from Ref.~\cite{Blanchard14}, were obtained from simulations with free and periodic boundary conditions, respectively. \\
     $^{d}$ The estimate $d_f = 1.65(3)$ was obtained directly from \eqref{eq-time-massscale} by using data at $t=1365\unit{s}$, which is however far from the interval where $M(t)$ is constant and far from the core of the $z = 2$ scaling regime. The other values $d_f = 1.613(14)$ and $1.609(9)$ are our final estimates obtained through the scaling relation \pref{eq-time-fractaldimension}. Please see \eqref{eq-time-df} and text for detail. \\
    \end{flushleft} 
    \end{table*}

  \begin{table*}[!t]
        \centering
        \caption{Universal scaling functions measured in the TNLC experiment and compared with theoretical predictions.}
        \label{table2}
        
        \begin{tabular}{lll}
        \hline\hline\noalign{\smallskip}
        Scaling function & Experimental result & Conclusion \\
        \noalign{\smallskip}\hline\noalign{\smallskip}
        Spatial correlator $\mathscr{F}(g)$ & \figref{Fig_Space-covariance-OJKTUG} & consistent with both OJK and TUG predictions \\
        Temporal correlator $\mathscr{H}(y)$ & \figref{Fig_Time-covariance}(b) & best described by LSI-2dIM prediction for $y \gtrsim 2$ \\
    \noalign{\smallskip}\hline\hline
    \end{tabular}
    \end{table*}
  
% --- Initial conditions; metaphor with Ising; asymmetry

  Because spatiotemporal correlations of the director field are negligible in the DSM2 state \cite{Joetz86}, DSM2 essencially generates random initial conditions. When electrically-switched ($t = 0 \unit{s}$) towards equilibrium, the ensuing competition between two possible conformations for the director field, in left- and right-handed twists (here noted $\phi = \pm 1$), resembles that between Ising spins when quenched from the higher-temperature ($T \gg T_c$, $T_c$ is the critical temperature) regime to the lower-temperature ($T \ll T_c$) phase [see \figref{Fig_Space-domains}]. Nonetheless, the experiment develops a weakly asymmetry likely along its dynamics: this is revealed by the ``magnetization'' $|\overline{M}'| \lesssim 0.2 \neq 0$, for $t \lesssim 60\unit{s}$ [\figref{Fig_Space-bubble}(a)], and by the shrinking rate of $\phi = \pm 1$ spherical domains (``bubbles'') [\figref{Fig_Space-bubble}(b)]. Noting the curvature-driven dynamics of disclinations (interfaces) [\eqref{eq-space-bubble}; \figref{Fig_Space-bubble}(b)], the asymmetry between the twisted phases also manifests in the Allen-Cahn diffusion coefficients, which take slightly different values ($4\%$ deviation from the mean $D_{\textrm{AC}} = 122(4) \unit{\mu m^2 s^{-1}}$) for $\phi = \pm 1$. As we discussed throughout the study, this weak asymmetry -- that is usual in experiments \cite{Yurke97,Sicilia_exp08} -- must be taken into account when one aims at testing theories for $\mathbb{Z}_2$-symmetric models.

% --- We have a Model A system 
    After observing that the experiment firmly shows well-known Model A features -- such as the dynamic exponent $z = 2$ [see \figref{Fig_Space-interface}; \eqref{eq-space-1z}] and dynamic scaling [\figref{Fig_Space-covariance}(b)] -- we move forward to comprehensively elucidate aspects that remain either controversial or not experimentally assessed yet for 2d Model A. Main results are summarized in Tables~\ref{table1} and \ref{table2}.

% --- Assess to Model A predictions not experimentally settled
    
    %--- What's new in space: BH prediction, OJK/TUG tests beyond Porod's regime (where it matters).
    %1
    From the spatial correlator at the Porod's regime, we measured the universal amplitude $A^*$ (see Table~\ref{table1}) in sharp agreement with the Bray-Humayun theory \cite{Bray_Humayun93_amplitudes}.
    %2
    Beyond the Porod's regime, the correlator form is well captured by both the Otha-Jasnow-Kawasaki (OJK) theory \cite{OJK82} and the theory of unstable growth (TUG) own to Mazenko \cite{Mazenko89,Mazenko90} [\figref{Fig_Space-covariance-OJKTUG}(b)]. Such conclusion emerges from comparisons done with no free parameters: the constant $D_\textrm{OJK}$ was found from its direct relation with $D_{\textrm{AC}}$ \cite{OJK82}, which we measured independently from the ``bubble'' experiments; $D_\textrm{TUG}$ was determined from the functional form in the Porod's regime. We have seen that although the agreement between experiment and theory is more pronounced for OJK (considering the most probable values for $D_\textrm{OJK}$ and $D_\textrm{TUG}$), TUG also may account for the TNLC correlator if the lower limit for $D_\textrm{TUG}$, set by its uncertainty, is used [\figref{Fig_Space-covariance-OJKTUG}(b)]. To our knowledge, this is the first experimental evidence\footnote{Interestingly, OJK-like form was also seen to describe the spatial correlations of  polycrystals formed by colloids and whose overall kinetics do not follow $z=2$ \cite{Lavergne17}. It is also the first time that the TUG two-point spatial correlator [see \eqref{eq-space-TUG-scalingfunctionF}] is confronted with experiment after Prof. Mazenko called for experimental tests in \cite{Mazenko91_2}. } that theoretical spatial correlators, such as OJK and TUG, describe experimental data in the nontrivial scale beyond the Porod's regime, for a curvature-driven ($z = 2$) system.

% --- What's new in time: 1) two-time correlator, aging, lambda, functional forms; 2) Persistence: DIF or 2dIM?
    
    %1
    Regarding the time correlations, we measured the two-time correlator after proposing its due modification to account for the weak asymmetry in the kinetics [\eqref{eq-time-cov-prime}]. These correlators, $C'_t(t,t_0)$, collapse on a master curve when rescaled by $y = t/t_0$, thus offering evidence of dynamic scaling hypothesis for 2d Model A systems as demanded by aging phenomenology \cite{Henkel} -- see \figref{Fig_Time-covariance}. The exponent $\lambda$ (\eqref{eq-time-covscaling}) displays a crossover from an initial OJK-like value to an asymptotic one which is compatible with both the FH conjecture and the TUG prediction (values in Table~\ref{table1}): this reinforces the difficult matter, witnessed by simulations, of distinguishing between the FH and TUG predictions. Light is shed by comparisons with the functional form itself, which elucidate that neither OJK or TUG theory can account for the experimental, 2d Model A, correlation function. Instead, the local scale invariance (LSI) theory own to Henkel \cite{Henkel02}, that describes the 2dIM correlator along with the FH conjecture \cite{Henkel04}, provides the best description for the experiment in the regime, $t/t_0 \gtrsim 2$, that LSI deals with [\figref{Fig_Time-covariance}(b)]. These results might indicate support to the FH conjecture from a global perspective, in harmony with many numerical studies \cite{Fisher88,Humayun91,Brown97,Henkel04,Corberi16,Christiansen20}. The agreement with LSI also raises up the question whether the paradigmatic 2d Model A systems may display conformal invariance in some stage of their kinetics \cite{Henkel02}.

    %2
    We have also assessed the local persistence statistics. Thanks to the accuracy of our data, we measure a local persistence exponent $\theta$ (\figref{Fig_Persistence-probability}; Table~\ref{table1}) in accordance with the recent estimates from 2dIM \cite{Blanchard14}, but in significant disagreement with the candidate 2d Model A value obtained from the diffusion equation \cite{Poplavskyi18} and the OJK theory \cite{Derrida96_OJK}. This experimentally-based observation thus points out the former as the genuine value in the 2d Model A class. Additionally, we have observed that the mosaic of persistence clusters, defined by the persistence index $\chi(\bm{r},t)$, develops a fractal morphology up to the scale of a correlation length $\zeta^*(t)\sim t^{1/2}$ (see \figref{Fig_Persistence-domains} and \figref{Fig_Persistence-massa}). The fractal dimension $d_f \approx 1.61$ of such morphology is measured experimentally (Table~\ref{table1}) and substantiated as universal \cite{Jain00}. Correlations among persistence clusters respect the dynamic scaling hypothesis [\figref{Fig_Persistence-massa}(e)].

 % --- Future: Different kinetics; thermal effects; percolation; response functions. 

    Having a versatile experimental model system in the Allen-Cahn (Model A) class opens novel paths for experimental investigation.
    The many interesting electroconvective patterns in the liquid crystals \cite{Joetz86, Joetz86_2, Kai89}, for instance, can be used to assess unusual sorts of ordering kinetics by transiting between nonequilibrium states, with the advantage of the genuinely sudden electrical switching. 
    Moreover, the independent control on the temperature allows one to tune thermal effects by adjusting the distance from the nematic-isotropic transition point \cite{DeGennes93}. Such control can be particularly important for experimentally assessing the fundamental theoretical hypothesis that, from the viewpoint of renormalization group \cite{AllenCahn79,Bray02,Arenzon07}, the bath temperature merely factorizes non-universal amplitudes, while it preserves universal features -- scaling exponents and correlation functions -- for a subcritical dynamics.
    Another appealing direction is suggested by the numerically observed \cite{Arenzon07} self-tuning of 2d Model A systems to the critical percolation fixed point \cite{Stauffer79,Araujo14,Saberi15}, which theoretically has boosted derivations of exact solutions for the distributions of hull areas \cite{Arenzon07,Sicilia_exp08}, perimeters \cite{Sicilia07}, domain crossing probabilities \cite{Olejarz12}, besides proposing a generalization of the dynamic scaling hypothesis \cite{Blanchardperc14,Corberi17}.
    Access to response functions \cite{Henkel02,Corberi16} is an additional and attractive avenue that calls for, and it is in principle amenable to, experimentation.

 \begin{acknowledgments}
 The authors acknowledge enlightening discussions with J. J. Arenzon, A. Azevedo-Lopes, F. Corberi, L. F. Cugliandolo, Y. T. Fukai, S. Majumdar, M. Mucciconi, M. Picco, and T. Ohta. This work was supported by Japan Society for the Promotion of Science (JSPS). RALA was funded by JSPS KAKENHI grant number JP16J06923. KAT was funded by JSPS KAKENHI grant numbers JP25103004, JP16H04033, JP19H05800, JP20H00128.
 \end{acknowledgments}

\bibliography{phaseordering.bib}

\end{document}